\documentclass[epj]{svjour}

\usepackage{amsmath}
\usepackage{amssymb}
\usepackage[cp1251]{inputenc}
\usepackage[T2A]{fontenc}
\usepackage{latexsym}
\usepackage{graphicx}
\usepackage[english]{babel}

\def\bge{\begin{equation}}
\def\ene{\end{equation}}
\def\bgea{\begin{eqnarray}}
\def\enea{\end{eqnarray}}

\begin{document}
\sloppy
\title{Translationally invariant calculations of form factors, nucleon densities and momentum distributions for finite nuclei with short-range correlations
included}
\author{A.~V.~ Shebeko\thanks{shebeko@kipt.kharkov.ua} \inst{1}
\and P.~A.~ Grigorov\thanks{grigorov@mail.ru}        \inst{2}
\and V.~S.~ Iurasov\thanks{iurasov90@gmail.com} \inst{3}
}
\institute{Institute for Theoretical Physics, NSC ``Kharkov Institute of Physics \& Technology'', Kharkiv 61108, Ukraine
\and
University of T\H{u}bingen, Germany
\and
National Karazin University, Ukraine}

\abstract{Relying upon our previous treatment of the density matrices
for nuclei (in general, nonrelativistic self-bound finite systems) we are studying a combined effect of
center-of-mass motion and short-range nucleon-nucleon correlations on the
nucleon density and momentum distributions in light nuclei ($^{4}He $ and $^{16}O$).
Their intrinsic ground-state wave functions are constructed in the so-called fixed center-of-mass approximation, starting
with mean-field Slater determinants modified by some correlator ({\it e.g.}, after
Jastrow or Villars). We develop the formalism based upon the Cartesian or boson representation,
in which the coordinate and momentum operators are linear combinations of the creation and annihilation operators
for oscillatory quanta in the three different space directions, and get the own "Tassie-Barker" factors
for each distribution and point out other model-independent results. After this separation of the center-of-mass motion effects
we propose additional analytic means in order to simplify the subsequent calculations ({\it e.g.}, within the Jastrow approach
or the unitary correlation operator method). The charge form factors,
densities and momentum distributions of $^{4}He $ and $^{16}O$ evaluated by using the well known cluster expansions
are compared with data, our exact (numerical) results and microscopic calculations.}

\maketitle

\section{Introduction}

Many efforts have been made to get a deeper understanding of the nuclear
structure at small distances (less than the pion Compton wavelength) with
realistic many-body calculations for the nuclear wave function (WF) whose
short-range part strongly deviates from a mean-field description. In this
respect, as well known (see, {\it e.g.}, survey \cite{Anton88}, ref. \cite{Alviol05} and refs. therein), the
nucleon density matrices and their Fourier transforms are of great interest,
being related, on the one hand, to the nuclear ground-state (g.s.)
properties and, on the other hand, to the cross sections of various medium-
and high-energy scattering processes off nuclei. Regarding the second aspect,
we mean firstly a comparatively simple relation in the Born approximation to
express the elastic electron scattering cross section through the charge
form factor (FF) $F_{ch}(q)$ of the target-nucleus and its charge density $%
\rho_{ch}(r)$ being defined by the Fourier transform of $F_{ch}(q)$. In
addition, in the so-called approximation of small interaction times (see
\cite{DOS76}-\cite{Dem88}) the double differential $(e,e^{\prime})$ reaction
cross section becomes proportional to an integral of the momentum
distribution (MD) $\eta(p)$ over the momentum range that is fixed with
certain combination (the $y-$scalling variable) of the momentum transfer $q$
and the energy transfer $\omega$ (cf. \cite{Ciofi85}). Other links
with $\eta(p)$ we find in approximate calculations of the spectral function
that determines the exclusive $A(e,e^{\prime}N)X$ cross sections (see, {\it e.g.},
review \cite{FruMo84}, ref. \cite{CioLiu91} and earlier papers \cite{Hofmann73},
\cite{JiKez75}).Of course, two-body and more complicated reaction mechanisms, in particular,
due to meson exchange currents (see, {\it e.g.}, \cite{KatAkaTan82} and \cite%
{SchiPaWi90}) in electromagnetic interactions with nuclei, may obscure such
links.

 Note also the distorted-wave-impulse-approximation calculations \cite{BeMeShe83}
of proton MDs in $^{12}C$ and $^{16}O(e, e^{\prime}p)$ reactions at Saclay kinematics,
where the authors have shown a strong enhancement of the reaction cross sections
with account for the final-state interaction at recoil momenta $q_R$ greater than
1.5 $fm^{-1}$. In the range the corresponding distributions of outgoing protons,
having a considerably slower fall-off with the $q_R$--increasing compared to
the plane-wave-impulse-approximation ones, may imitate some SRC effect.
Therefore, the corresponding theoretical approaches are needed in
certain refinements to bringing a reliable information on the distributions
in question from experimental data. Neglecting these complexities one has to
deal \cite{DOS76}, \cite{KorShe77}, \cite{ShePaMav06} with the two structure
quantities, {\it viz.}, the intrinsic density distribution (DD) or simply the
intrinsic density $\rho_{int}(r)$ and the intrinsic MD $\eta_{int}(p)$. They
are expectation values in the translationally invariant (intrinsic) g.s. WF
of appropriate many-body (multiplicative) operators which depend on the
respective Jacobi variables. These definitions (see the next section)
coincide with those by the Sapporo group \cite{Morita87}, \cite{Morita88} in
studying the properties of few-body systems, but differ from the ones used
by the authors of refs. \cite{Alviol05}, \cite{Massen99}, \cite{Massen00} in
their calculations of the densities and momentum distributions in $s-p$ and $%
s-d$ shell nuclei. There we encounter the other (not intrinsic) quantities $%
\rho(r)$ and $n(k)$ introduced as in the case of infinite systems ({\it e.g.}, the
nuclear matter) by means of the expectation value of the one-body "density
operator" with a trial Jastrow-type WF. The latter in its schematic form $%
\Psi=\hat{F}\Phi$ involves a correlation operator $\hat{F}$\footnote{%
Below, the notation $\hat F = \hat C$ is employed as well}, which
incorporates correlations into the mean-field WF $\Phi$. It is required that
$\hat{F}$ be translationally invariant and symmetrical in particle
permutations. However, when starting with a Slater determinant (SD) $\Phi$,
{\it e.g.}, as in \cite{Alviol05},\cite{Massen99} the function $\Psi$ is
translationally non-invariant ("bad"), that is, it contains spurious
components which result from the CM motion (CMM) in a non-free state. In this
connection, let us recall earlier and more recent attempts \cite{Fria71}-%
\cite{Navrat04} to remedy such a deficiency of the nuclear WF, namely its
lack of translational invariance (TI) wherever shell-model WFs (commonly
built up from single-particle (s.p.) orbitals) are used.

In most cases the CM correction has been made to calculate the FF $F_{ch}(q)$
and, respectively, the density $\rho_{ch}(r)$ using, as a rule, the
Tassie-Barker (TB) prescription (a comparison of the relevant effects can be
found in ref. \cite{ShePaMav06}) while the not intrinsic DM $n(k)$ has been
corrected (without any good reasons) via the renormalization $%
b\longrightarrow\sqrt{\frac{A-1}{A}}b$ of the corresponding oscillator
parameter $b$ (see, {\it e.g.}, \cite{YpGr95}), {\it i.e.}, as in the case of $%
\rho_{ch}(r)$. An alternative evaluation \cite{DOS76}-\cite{Dem88}, \cite%
{ShePaMav06} of the intrinsic FF's, densities and momentum distributions,
put forward in \cite{DOS76} to overcome some obstacles in describing the
elastic and inclusive electron scattering off the $^4He$ nucleus, has
brought a fresh look at the CM correction of these quantities. In
particular, it turns out that $\rho_{int}(r)$ and $\eta_{int}(p)$ are shrunk
(from the periphery of each of them to its central part) compared to $%
\rho(r) $ and $\eta(p)$. To our knowledge, this significant consequence of
the restoration of TI has been ignored in past and goes on to be missed in
modern explorations \cite{Alviol05}, \cite{Massen00}.

At this point, one should note that such a simultaneous shrinking of the
density and momentum distributions has been found within the harmonic
oscillator model (HOM) for the simple $(1s)^{4}$ configuration. Accordingly,
the motivation of the present work is twofold. First, we will show our
results obtained with WFs more realistic than the $1s-shell$ SD composed of
harmonic oscillator (HO) orbitals. Second, the approach of \cite{DOS76},\cite%
{ShePaMav06} is extended to heavier nuclei (cf. \cite{KorShe77} ). The CM
correction of their FFs and MDs is considered on an equal physical footing,
{\it viz.}, using one and the same translationally g.s. WF that incorporates the
nucleon-nucleon short-range correlations (SRCs).

We employ  Jastrow WFs  \cite{Jast55} and the unitary-model-operator
($\exp(\imath S)$ with ${S}^{\dagger} = S$) approach (UCOA)
\cite{Villa63}, \cite{ProviShak64} and \cite{ShaWagHu67} to nuclear-structure physics
and its development by the Darmstadt group \cite{Feld98}, \cite{RotNefFe2010}
(cf. the diagram-free (coupled-channel) $\exp(S) $ - method with ${S}^{\dagger} \neq -S$ in
the many-fermion theory \cite{KumLuZab78}). In the context, let us remind other methods of
deriving the so-called cluster expansions for the expectation values with respect to
Jastrow WFs \cite{IwaYam57}- \cite{DalStringBoh82}.
Among them we note a factor-cluster or Van Kampen-type expansion  proposed in \cite{ClarkRist70} to evaluate
the distributions of interest with special emphasis upon the correlated charge FF for elastic electron
scattering off nuclei. It has turned out that the expansion is equivalent to an approximate version of the UCOA,
described in \cite{CiofiGryp69}, and yields a factor-cluster analogue of the Iwamoto-Yamada expansion \cite{IwaYam57}.
The former (called sometimes the FIY expansion) simplifies numerical calculations compared to the latter.
A careful comparison of the correlated one-body properties of $s-p$ and $s-d$ nuclei, evaluated within the Jastrow formalism
by truncating
the FIY, FAHT ( factor analogue of the expansion from \cite{Aviles58}-\cite{HartTolhoek58}) and in the low-order
approximation (LOA) from \cite{GaudGillRipka71} for the one-body density matrix (1DM), has been carried out
in \cite{MoustMassPanGrypAnt01}.
In the three cases the CMM correction has been taken into
account by the commonplace TB factor when extracting the model parameters (the HO parameter and correlation radius ) from the experimental charge FF (we will come
back to the point later). Of great interest are also the exact Jastrow calculations  of the elastic FF, MD
and two-body density of $^{4}He$ performed in \cite{DalStringBoh82} without any CMM correction (see our discussion below).

The paper is organized as follows. The underlying formalism with basic definitions is exposed in
in the following section. Sect. 3 is devoted to constructing the translationally invariant
correlated WFs, while sect. 4 is contained the formulae obtained with the help of the UCOA decomposition of
the similarity transformation ${\hat C}^{\dagger} {\hat O}^{[1]}\hat C$ truncated at the two-body terms. Here ${\hat O}^{[1]}$
is a relevant one-body operator additive by nucleons. Explicit expressions for the  DDs and MDs of nucleons in
$^{4}He $ and $^{16}O$ are shown together with their FFs separately in subsect. 4.1 and 4.2. Our results are
discussed and compared with the data in sect. 5. Some intermediate derivations can be found in Appendices.

\section{The intrinsic form factor, density and momentum distributions and
their evaluation in the Cartesian representation}
By definition, the intrinsic (elastic) FF of a nonrelativistic system with
the mass number $A$ and the total angular momentum equal to zero is
\begin{equation}  \label{e:ff1}
F(q)=F_{int} (q) \equiv \frac 1A \sum\limits_{\alpha=1}^A \langle \Psi_{int}
\mid \exp[\imath \vec {q} \cdot (\hat {\vec {r}}_\alpha - \hat {\vec {R}})]
\mid \Psi_{int} \rangle
\end{equation}
or
\[
F(q)=\langle \Psi_{int} \mid \exp[\imath \vec {q} \cdot (\hat {\vec {r}}_1 -
\hat {\vec {R}})] \mid \Psi_{int} \rangle=\dots
\]
\[
=\langle \Psi_{int} \mid \exp[\imath \vec {q} \cdot (\hat {\vec {r}}_A -
\hat {\vec {R}})] \mid \Psi_{int} \rangle,
\]
where $\Psi_{int} $ is the intrinsic WF of the system (nucleus), $\hat {%
\vec {r}}_\alpha$ the coordinate operator for nucleon number $\alpha$, and $%
\hat {\vec {R}} = A^{-1} \sum_{\alpha=1}^A \hat {\vec {r}}_\alpha$ the CM
operator.

Recall that $\mid \Psi _{int}\rangle $ enters the eigenvector $|\Psi _{\vec{P%
}}\rangle $ of the total Hamiltonian $\hat{H}$ of the system, which belongs
to the eigenvalue $\vec{P}$ of the total momentum operator $\hat{\vec{P}}%
=\sum\limits_{\alpha =1}^{A}\hat{\vec{p}}_{\alpha }$:
\begin{equation}
|\Psi _{\vec{P}}\rangle =|\vec{P})\ |\Psi _{\mathrm{int}}\rangle .
\end{equation}%
Here $\vec{p}_{\alpha }$ is the momentum operator of the $\alpha $-th
particle. Henceforth the bracket $|\,\,)$ is used to represent a vector in
the space of the center-of-mass coordinates, so that $\hat{\vec{P}}|\vec{P})=%
\vec{P}|\vec{P})$. A ket~(bra) with an index $|\cdots \rangle _{\alpha }$~($%
_{\alpha }\langle \cdots |$) will refer to the state of the $\alpha -$th
particle. The intrinsic WF $\Psi _{\mathrm{int}}$ depends upon the $A-1$
independent intrinsic variables. These may be expressed in terms of the
Jacobi coordinates, {\it e.g.},
\begin{equation}
{{\vec{\xi}}}_{\alpha }={{\vec{r}}}_{\alpha +1}-\frac{1}{{\alpha }}%
\sum_{\beta =1}^{\alpha }{{\vec{r}}}_{\beta }\qquad (\alpha =1,2,\ldots
,A-1)\,  \label{e:ja}
\end{equation}%
or the corresponding canonically conjugate momenta
\begin{equation}
{{\vec{\eta}}}_{\alpha }=\frac{1}{\alpha +1}(\alpha {{\vec{p}}}_{\alpha
+1}-\sum_{\beta =1}^{\alpha }{{\vec{p}}}_{\beta })\qquad (\alpha =1,2,\ldots
,A-1).
\end{equation}%
The WF $\Psi _{\vec{P}}(\vec{r}_{1},\vec{r}_{2},...,\vec{r}_{A})$ in the
coordinate representation satisfies the requirement of TI,
\begin{equation}
\Psi _{\vec{P}}(\vec{r}_{1}+\vec{a},\vec{r}_{2}+\vec{a},\ldots ,\vec{r}_{A}+%
\vec{a})=\exp (\mathrm{i}\vec{P}\cdot \vec{a})\Psi _{\vec{P}}(\vec{r}_{1},%
\vec{r}_{2},...,\vec{r}_{A}),  \label{Etrinv}
\end{equation}%
for any arbitrary displacement $\vec{a}$.

The intrinsic density $\rho_{\mathrm{int}}(r)$ is the Fourier transform of the
elastic FF, or inversely,
\begin{equation}  \label{e:fff}
F_{\mathrm{int}}(\vec{q}) = \frac{1}{A}\int \mbox{e}^{\mathrm{i} \vec{q}\cdot%
\vec{r} } \rho_{\mathrm{int}} (\vec{r}) \mbox{d}^3r .
\end{equation}
From eq.(\ref{e:fff}) it follows that $\rho_{\mathrm{int}} (\vec{r}) = A
\langle \Psi_{\mathrm{int}}| \hat{\rho}_{\mathrm{int}} (\vec{r}) | \Psi_{%
\mathrm{int}} \rangle $, where
\begin{equation}  \label{e:rhor}
\hat{\rho}_{\mathrm{int}} (\vec{r}) = \delta ( \vec{r} - \hat{\vec{r}}_A +
\hat{\vec{R}} ) = \delta ( \vec{r} - \mbox{$\frac{A-1}{A}$}\hat{\vec{\xi}}%
_{A-1}).
\end{equation}

Further, the 1DM may be defined as
\begin{eqnarray}
\rho^{[1]}_{\mathrm{int}} (\vec{r},\vec{r^{\prime}}) & \equiv & A
\langle \Psi_{\mathrm{int}} | \hat{\rho}_{\mathrm{int}}^{[1]} (\vec{r},\vec{%
r^{\prime}}) | \Psi_{\mathrm{int}} \rangle \nonumber\\
&=& A\langle \Psi_{\mathrm{int}} |{\ \vec{\xi}}_{A-1} = \vec{r} \rangle
\langle {\vec{\xi}}_{A-1} = \vec{r^{\prime}} | \Psi_{\mathrm{int}} \rangle \nonumber\\
&=&\! A \int \!\! \mbox{d}^3\xi_1\ldots\mbox{d}^3\xi_{A-2} \Psi^{\dagger}_{%
\mathrm{int}}(\vec{\xi}_1,\ldots , \vec{\xi}_{A-2},\vec{r})  \nonumber \\
& & \times \Psi_{\mathrm{int}}(\vec{\xi}_1,\ldots , \vec{\xi}_{A-2},\vec{%
r^{\prime}}) ,
\label{e:ROINT2}
\end{eqnarray}

so that the normalization condition $\int\mbox{d}^3 r \rho^{[1]}_{\mathrm{int%
}} (\vec{r},\vec{r}) = A$ is satisfied. We would like to emphasize that this
is not an ``imposed" definition. It appears naturally when evaluating the
dynamical FF \cite{GonShe74} (or its diagonal part, if one uses the
terminology adopted in Chapter XI of the monograph \cite{GW64}), which is
related to the intrinsic MD \cite{KorShe85}
\begin{equation}  \label{e:np1}
\eta_{\mathrm{int}} (\vec{p}) \equiv A \langle \Psi_{\mathrm{int}} | \hat{%
\eta}_{\mathrm{int}}(\vec{p}) | \Psi_{\mathrm{int}}\rangle
\end{equation}
with
\begin{eqnarray}
\hat{\eta}_{\mathrm{int}}(\vec{p}) &=& \delta (\vec{p}-\hat{\vec{p}}_A +
\hat{\vec{P}}/A) = \delta (\vec{p}- \hat{\vec{\eta}}_{A-1} )  \label{e:npo}
\nonumber\\
&=& | {\vec{\eta}}_{A-1} = \vec{p}\rangle \langle {\vec{\eta}}_{A-1}=\vec{p}
| .
\end{eqnarray}

The OBMD is the Fourier transform of the 1DM $\rho^{[1]}_{\mathrm{int}} (%
\vec{r},\vec{r^{\prime}}) $,
\begin{equation}  \label{e:npf}
\eta_{\mathrm{int}} (\vec{p}) = (2\pi )^{-3} \int \mbox{d}^3r\mbox{d}%
^3r^{\prime}\exp{[\mathrm{i}\vec{p}\cdot (\vec{r}-\vec{r^{\prime}})]} \rho_{%
\mathrm{int}}^{[1]}(\vec{r},\vec{r^{\prime}}) .
\end{equation}

As in \cite{ShePaMav06} we would like to point out that
\begin{equation}  \label{e:rhorel}
\rho_{\mathrm{int}} (\vec{r}) = \left[\mbox{$\frac{A}{A-1}$}\right]^3
\rho^{[1]}_{\mathrm{int}} (\mbox{$\frac{A}{A-1}$}\vec{r},\mbox{$%
\frac{A}{A-1}$}\vec{r}) .
\end{equation}
In other words, the intrinsic 1DM does not have the property $\rho^{[1]}(%
\vec{r}) = \rho^{[1]}(\vec{r},\vec{r})$ which can be justified for infinite
systems, although it has often been exploited in approximate treatments of
finite systems (cf., however, ref.~\cite{VNW1998}, where an alternative
definition of the 1DM for finite self-bound systems was proposed).

Each of these intrinsic quantities can be written as the expectation value
of a product of $A$ operators acting on the subspaces of the separate $A$
particles. For example, we have
\begin{equation}
F_{int}(q)\equiv\langle \Psi_{int} \mid \hat{F}_{int}(\vec
q)|\Psi_{int}\rangle
\end{equation}
with the multiplicative operator
\[
\hat{F}_{\mathrm{int}} (\vec{q}) = \exp [ \mathrm{i}\vec{q} \cdot (\hat{\vec{%
r}}_1 - \hat{\vec{R}} )] = \mbox{e}^{ \mathrm{i} \frac{A-1}{A} \hat{\vec{r}}%
_{1} \cdot \vec{q}}\mbox{e}^{-\mathrm{i} \frac{\hat{\vec{r}}_2 }{A}\cdot\vec{%
q}}\ldots \mbox{e}^{-\mathrm{i} \frac{\hat{\vec{r}}_A }{A}\cdot\vec{q}},
\]
whereas
\[
{\hat {\rho}}_{\mathrm{int}} (\vec{r}) = \delta ( \hat{\vec{r}}_1 - \hat{%
\vec{R}} - \vec{r} ) = (2\pi)^{-3}\int \mbox{e}^{- \mathrm{i} \vec{q}\cdot
\vec{r} } \hat{F}_{\mathrm{int}} (\vec{q}) \mbox{d}^3q .
\]

Now, we will use the Cartesian representation, in which the coordinate
\thinspace (momentum) operator $\hat{\vec{r}}_{\alpha }$ ($\hat{\vec{p}}%
_{\alpha }$) of the $\alpha $-th particle is the linear combination of the
Cartesian creation and annihilation operators ${\hat{\vec{a}}}^{\dag }$ and $%
\hat{\vec{a}}$ ,
\begin{equation}
\hat{\vec{r}}=\frac{r_{0}}{\sqrt{2}}({\hat{\vec{a}}}^{\dag }+{\hat{\vec{a}}}%
),\hspace{4mm}\hat{\vec{p}}=\mathrm{i}\frac{p_{0}}{\sqrt{2}}({\hat{\vec{a}}}%
^{\dag }-{\hat{\vec{a}}}),\hspace{4mm}r_{0}p_{0}=1,  \label{e:cao}
\end{equation}%
with the Bose commutation rules
\begin{eqnarray}
\lbrack \hat{a}_{l}^{\dag },\hat{a}_{j}^{\dag }]=[\hat{a}_{l},\hat{a}_{j}]=0%
\hspace{5mm},\hspace{5mm}[\hat{a}_{l},\hat{a}_{j}^{\dag }]=\delta _{lj}.  \label{e:com1}
\end{eqnarray}%
The indices $l,j=1,2,3$ label the three Cartesian axes $x,y,z$.

As the "length parameter" $r_0$ one can choose the oscillator parameter of a
suitable HO basis in which the nuclear WF is expanded. Its basis vectors $%
|n_x\,n_y\,n_z\rangle_1 \otimes \ldots \otimes |n_x\,n_y\,n_z\rangle_A $,
where the quantum numbers $n_x,\,n_y,\,n_z$ take on the values $0,1,\ldots ,$
are composed of the s.p. states
\begin{equation}  \label{e:bas}
|n_x\,n_y\,n_z\rangle = \left[n_x !\,n_y !\,n_z ! \right]^{-\frac12} \left[
\hat{a}_1^\dag \right]^{n_x} \,\left[ \hat{a}_2^\dag \right]^{n_y} \, \left[
\hat{a}_3^\dag \right]^{n_z} | 0\,0\,0 \rangle \, ,
\end{equation}
which are the eigenstates of the Hamiltonian $\hat{H}_{\mathrm{osc}} =
\omega ( \hat{\vec{a}}^{+} \cdot \hat{\vec{a}} + \frac32 ) $,
\[
\hat{H}_{\mathrm{osc}} |n_x\,n_y\,n_z\rangle = ( n_x + n_y + n_z + %
\mbox{$\frac32$} )\, \omega \, |n_x\,n_y\,n_z\rangle \ ,
\]
where $\omega$ is the oscillation frequency along the three axes $x,y$ and $%
z $. {We use the system of units with $\hbar=c=1$}. The s.p. WF in
coordinate representation is written
\[
\langle \vec{r} \mid n_x\,n_y\,n_z\rangle = \psi_{n_x} (x) \psi_{n_y} (y)
\psi_{n_z} (z) \, ,
\]
where (see, {\it e.g.}, \cite{NS69})
\[
\psi_{n} (s) = \left[ \sqrt{\pi} 2^n n! r_0 \right]^{- \frac12} H_n ( {s /
r_0} ) \exp (- {s^2 / 2 r_0^2} )
\]
and $H_n(x)$ is a Hermite polynomial. By definition, the oscillator
parameter equals $r_0 = [m \omega]^{-\frac12} $.

Using eqs. (\ref{e:cao}- \ref{e:com1}), after some algebra one can get
\[
\hat{F}_{int}(\vec{q}) =F_{TB}(q)~F_{HOM}(q)\times
\]
\[
\times\exp{\left[\imath\vec{q}\left(\frac{A-1}{A}\right)\frac{r_0}{\sqrt{2}}~%
\hat{\vec{a}}_1^{\dag}\right]}\times
\]
\[
\times\exp{\left[\imath\vec{q}\left(\frac{A-1}{A}\right)\frac{r_0}{\sqrt{2}}~%
\hat{\vec{a}}_1\right]}\times
\]
\[
\times\exp{\left[-\imath\vec{q}\frac{r_0}{\sqrt{2}A}~\hat{\vec{a}}_2^{\dag}%
\right]} \exp{\left[-\imath\vec{q}\frac{r_0}{\sqrt{2}A}~\hat{\vec{a}}_2%
\right]}\dots
\]
\begin{equation}
\times\exp{\left[-\imath\vec{q}\frac{r_0}{\sqrt{2}A}~\hat{\vec{a}}_A^{\dag}%
\right]} \exp{\left[-\imath\vec{q}\frac{r_0}{\sqrt{2}A}~\hat{\vec{a}}_A%
\right]},\label{20}
\end{equation}
with $F_{TB}(q)=~\exp(\frac{1}{4A}q^2 r_0^2)$,\,~$F_{HOM}(q)~=~\exp(-\frac{
1}{4}q^2 r_0^2)$.\\

Thereat, the TB factor $F_{TB}(q)$ appears automatically due
to a specific structure of the operators involved. In other words, its
appearance is independent of any nuclear properties (in general, properties
of a finite system). The only mathematical tool that has been used is the
Baker-Hausdorff relation:
\begin{equation}
\mathrm{{e}^{\hat{A}+\hat{B}}={e}^{\hat{A}}~{e}^{\hat{B}}~{e}^{-\frac{1}{2}[%
\hat{A},\hat{B}]},}  \label{BakHaus}
\end{equation}%
that is valid with arbitrary operators $\hat{A}$ and $\hat{B}$ for which the
commutator $\left[ \hat{A},\hat{B}\right] $ commutes with each of them.

\subsection{Constructing intrinsic wave functions. Inclusion of nucleon-nucleon
correlations}

A Slater determinant,
\begin{equation}
\mid Det \rangle=\frac{1}{\sqrt{A!}}\sum_{\mathit{\hat{\mathcal{P}} \in S_A}%
} \epsilon_{ \mathcal{P}} \hat{\mathcal{P}}\{\mid \phi_{p_1}(1)\rangle \dots
\mid \phi_{p_A}(A)\rangle\},\label{22}
\end{equation}
as the total WF $\Phi$ for an approximate and convenient description of the
nuclear g.s., in the framework of the IPM or the Hartree-Fock(HF) approach
exemplifies WF's which do not possess the property of TI, eq.(\ref{Etrinv}).
Here $\epsilon_{ \mathcal{P}}$ is the parity factor for the permutation $%
\mathcal{P}$, $\phi_a$ the occupied orbital with the quantum numbers $\{a\}$
and the summation runs over all permutations of the symmetric group $S_A$.

There are different ways to restore TI if one starts with such a bad WF as $%
\mid Det\rangle$ (\cite{PeY1957}-\cite{ScG1990a}, \cite{Schmid0103}).

According to Ernst, Shakin and Thaler (EST) prescription \cite{EST1973}
\footnote{%
Other projection recipes can be applied without essential changes, see \cite%
{ShePaMav06}} in the fixed-CM approximation the nuclear many-body WF with
the total momentum $\vec{P}$ can be written in the form:
\begin{equation}
\mid \Psi_{P} \rangle =|\vec{ P}) \mid \Psi^{EST}_{int} \rangle.\label{23}
\end{equation}
The intrinsic WF after EST
\begin{equation}
\mid \Psi^{EST}_{int} \rangle = \frac{ (\vec{R}=0 \mid \Phi \rangle} {%
[\langle \Phi \mid \vec{R}=0)(\vec{R}=0 \mid \Phi \rangle]^{1/2}}\label{24}
\end{equation}
is constructed from an arbitrary (in general, translationally non-invariant)
WF $\Phi$, by requiring that the CM coordinate $\vec {R}$ be equal to zero.
The corresponding FF is the ratio
\[
F_{EST}(q) = \frac{A(q)}{A(0)},
\]
\begin{equation}
A(q)= \langle \Phi \mid {(2\pi)}^3 \delta (\hat {\vec {R}}) \exp[i \vec {q}
\cdot (\hat {\vec {r}}_1 - \hat {\vec {R}})] \mid \Phi \rangle,
\label{25}
\end{equation}
while the intrinsic MD
\begin{equation}
\eta_{EST}(p) = \frac{\langle \Phi \mid {(2\pi)}^3 \delta (\hat {\vec {R}})
\delta (\hat {\vec {p}}_1 - \hat {\vec {P}}/A - \vec{p}) \mid \Phi \rangle}{%
\langle \Phi \mid {(2\pi)}^3 \delta (\hat {\vec {R}}) \mid \Phi \rangle}
\label{26}
\end{equation}
so that we have the Fourier transform
\begin{equation}
\eta_{EST}(p)=(2\pi)^{-3}\int\exp(-\imath\vec{p}\vec{z})N(z)/N(0)d\vec{z}
\label{27}
\end{equation}
with
\begin{equation}
N(z)=\langle \Phi\mid (2\pi)^{3} \delta(\vec{R})\exp[\imath(\vec{p}_1-\vec{P}%
/A)\vec{z}]\mid\Phi\rangle.
\label{28}
\end{equation}
We see the certain resemblance between the structure functions $N(z)$ and $%
A(q)$, \emph{viz.}, both are determined by the expectation values of similar
multiplicative operators with one and the same trial WF $\Phi$. Owing to
this with the help of the same algebraic techniques (cf. eq. (\ref{20}) ) we get
\begin{equation}
A(q)=\exp{\left(-\frac{ q^{2} \bar{r}_0^2}{4} \right)}U(q),\label{29}
\end{equation}
\begin{equation}
U(q)=\int{d\vec{\lambda}\exp{\left(-\frac{r_0^2\lambda^2}{4A}\right)} F(\vec{%
v},\vec{s})},\label{30}
\end{equation}
with
\begin{equation}
\vec{s}=\imath\frac{r_0}{\sqrt{2}}~\vec{q}, \qquad \vec{v}=\imath\frac{r_0}{%
\sqrt{2}A}(\vec{\lambda}-\vec{q})\label{31}
\end{equation}
and the renormalized "length" parameter
\[
\bar{r}_0 = \sqrt{\frac{A-1}{A}}r_0
\]
and, in parallel,
\begin{equation}
N(z)=\exp\left( - \frac{z^2 \bar{p}_0^2}{4} \right) D(z),\label{32}
\end{equation}
\begin{equation}
D(z)=\int d \vec{\lambda} \exp\left(-\frac{r_0^2\lambda^2}{4A} \right)F(\vec{%
v}~^{\prime},\vec{s}~^{\prime}),\label{33}
\end{equation}
with
\begin{equation}
\vec{s}~^{\prime}=-\frac{p_0}{\sqrt{2}}~\vec{z}, \qquad \vec{v}~^{\prime}=%
\frac{\imath r_0}{\sqrt{2}A}(\vec{\lambda}-\imath p_0^2\vec{z})\label{34}
\end{equation}
and
\[
\bar{p}_0 = \sqrt{\frac{A-1}{A}}p_0.
\]
When deriving these relations we have applied again eq.(\ref{BakHaus}) in
combination with the representation
\begin{equation}  \label{delta}
\left(2\pi\right)^3\delta\left(\hat{\vec{R}}\right) = \int{\exp{\left(\imath%
\vec{\lambda}\hat{\vec{R}}\right)d\vec{\lambda}}}.
\end{equation}
After this we see that the expectations $A(q)$ and $N(z)$ are expressed
through one and the same function $F(\vec x, \vec y)$
\begin{equation}
F(\vec{x},\vec{y})=\langle\Phi|\hat{O}_1(\vec{x}+\vec{y}~)\hat{O}_2(\vec{x}%
)\dots\hat{O}_A(\vec{x})|\Phi\rangle,\label{36}
\end{equation}
where
\begin{equation}
\hat{O}_\gamma(\vec{x})=\exp(-\vec{x}^\ast\hat{\vec{a}}^\dag_\gamma)\exp(%
\vec{x}\hat{\vec{a}}_\gamma) \equiv\hat{E}^\dag_\gamma(-\vec{x})\hat{E}%
_\gamma(\vec{x})\label{37}
\end{equation}
\[
(\gamma=1,\dots,A).
\]
In other words, we have constructed the generating function for both. One
should stress that this result has been obtained independently of the model
WF $\Phi$.

Following a common practice  let us consider a correlated
A-body trial WF,
\begin{equation}
\mid \Phi \rangle=\mid \Phi_{corr} \rangle =\hat{C}(1,2, \cdots, A )\mid Det
\rangle.\label{38}
\end{equation}
The \emph{A}-particle operator $\hat{C} = C(\hat{\vec{r}}_\alpha-\hat{\vec{r}%
}_\beta, ~\hat{\vec{p}}_\alpha-\hat{\vec{p}}_\beta) $ \footnote{%
Of course, the operator may be spin and isospin dependent} introduces the
SRCs and meets all necessary requirements of the translational and Galileo
invariance, the permutable and rotational symmetry, {\it etc}. However, being
translationally invariant itself such a model introduction of correlations
does not enable to restore the TI violated with such a shell-model WF as the
Slater determinant.

What follows can be used with the Jastrow correlator \cite{Jast55}
\begin{equation}
\hat{C}=\frac{\hat{J}}{\sqrt{C_{J}}},\qquad \hat{J}=\prod\limits_{\alpha
<\beta }^{A}f(\hat{\vec{r}}_{\alpha \beta })\label{39}
\end{equation}%
The normalization constant $C_{J}=\langle Det\mid J^{\dag }J\mid Det\rangle $
(in general, a constant $\langle Det\mid C^{\dag }C\mid Det\rangle $, if any
) may be omitted keeping in mind the ratios $A(q)/A(0)$ and $N(z)/N(0)$. The
function $f(r_{\alpha \beta })$ of the distance $r_{\alpha \beta }=|{\vec{r}}%
_{\alpha }-{\vec{r}}_{\beta }|$ is required to come to zero when particles $%
\alpha $ and $\beta $ are inside a correlation volume of a radius $r_{c}$.

Another popular option goes back to the lectures by Villars in \cite{Villa63}
(see also \cite{ProviShak64}) with a unitary operator
\begin{equation}
\hat{C}=\exp(-\imath \hat{G}),\label{40}
\end{equation}
\begin{equation}
\hat{G}=\sum\limits_{\alpha<\beta}\hat{g}(\alpha,\beta),\label{41}
\end{equation}
where the Hermitian operator $\hat{g}(\alpha,\beta)$ acts onto the space of
the pair $(\alpha,\beta)$. In particular, we could follow the simplest
Darmstadt ansatz \cite{Feld98}:
\begin{equation}
\hat{g}(\alpha,\beta) =\frac{1}{2}\{\vec{s}~(\hat{\vec{r}}_{\alpha\beta})%
\hat{\vec{p}}_{\alpha\beta}+ \hat{\vec{p}}_{\alpha\beta}\vec{s}~(\hat{\vec{r}%
}_{\alpha\beta}) \},\label{42}
\end{equation}
where $\vec{s}$ is a function of the relative coordinate $\hat{\vec{r}}%
_{\alpha\beta}=\hat{\vec{r}}_\alpha-\hat{\vec{r}}_\beta$. Its canonically
conjugate momentum $\hat{\vec{p}}_{\alpha\beta}=\frac{1}{2}(\hat{\vec{p}}%
_\alpha-\hat{\vec{p}}_\beta)$.

Keeping in mind similar constructions we rewrite expectation (\ref{36}) as
\begin{equation}
F(\vec{x},\vec{y})=\langle\Phi(-\vec{x})\mid \hat{E}^\dag_1(-\vec{y})\hat{E}%
_1(\vec{y})\mid\Phi(\vec{x})\rangle,\label{43}
\end{equation}
where
\[
\mid\Phi(\vec{x})\rangle=\hat{E}_1(\vec{x})\dots\hat{E}_A(\vec{x}%
)\mid\Phi\rangle,
\]
since $\hat{E}_1(\vec{x}+\vec{y})=\hat{E}_1(\vec{x})\hat{E}_1(\vec{y})$ and $%
[\hat{E}_\alpha(\vec{x}),\hat{E}_\beta(\vec{y})]=0~~\newline
(\alpha,\beta=1,\dots,A)$ for any vectors $\vec{x}$ and $\vec{y}$.

Moreover, we find that
\begin{equation}
\hat{E}(\vec{x})~\hat{\vec{r}}~\hat{E}^{-1}(\vec{x})=\hat{\vec{r}}+\frac{r_0%
}{\sqrt{2}}~\vec{x}\label{44}
\end{equation}
and
\begin{equation}
\hat{E}(\vec{x})\hat{\vec{p}}~\hat{E}^{-1}(\vec{x})=\hat{\vec{p}}-\imath%
\frac{p_0}{\sqrt{2}}~\vec{x}.\label{45}
\end{equation}
Remind that $E^\dag\neq E^{-1}$. In other words, $\hat{E}_\alpha(\vec{x})$
is the displacement operator in the space of nucleon states with the label $%
\alpha$.

Due to this property when handling the similarity transformation
\[
\hat{C}^{\prime}=\hat{E}_1(\vec{x})\dots\hat{E}_A(\vec{x}) C(\hat{\vec{r}}%
_\alpha-\hat{\vec{r}}_\beta,~\hat{\vec{p}}_\alpha- \hat{\vec{p}}%
_\beta)\times
\]
\[
\times\hat{E}^{-1}_1(\vec{x})\dots\hat{E}^{-1}_A(\vec{x}),
\]
we get
\[
\hat{C}^{\prime}=C(\hat{E}_\alpha(\vec{x})\hat{\vec{r}}_\alpha\hat{E}%
^{-1}_\alpha(\vec{x})- \hat{E}_\beta(\vec{x})\hat{\vec{r}}_\beta\hat{E}%
^{-1}_\beta(\vec{x}),
\]
\[
\hat{E}_\alpha(\vec{x})\hat{\vec{p}}_\alpha\hat{E}^{-1}_\alpha(\vec{x})-
\hat{E}_\beta(\vec{x})\hat{\vec{p}}_\beta\hat{E}^{-1}_\beta(\vec{x}))=
\]
\[
=C(\hat{\vec{r}}_\alpha-\hat{\vec{r}}_\beta,~\hat{\vec{p}}_\alpha- \hat{\vec{%
p}}_\beta)=\hat{C}
\]
\emph{i.e.},
\begin{equation}
\hat{C}^{\prime}=\hat{C}.\label{46}
\end{equation}
Recall that $C$ is a function of \underline{all} the relative coordinates
and their canonically conjugate momenta.\newline
From eqs. (\ref{38}) and (\ref{46}) it follows that
\[
\mid \Phi_{corr}(\vec{x})\rangle \equiv \hat{E}_1(\vec{x})\dots\hat{E}_A(%
\vec{x}) \mid \Phi_{corr} \rangle=
\]
\begin{equation}
=\hat{C}\mid Det(\vec{x})\rangle.\label{47}
\end{equation}
Here $\mid Det(\vec{x})\rangle= \hat{E}_1(\vec{x})\dots\hat{E}_A(\vec{x}%
)\mid Det \rangle$ is a new Slater determinant composed of the renormalized
orbitals,
\begin{equation}
\mid \phi_{a}(\alpha; \vec{x})\rangle= \hat{E}_{\alpha}(\vec{x}%
)\mid\phi_{a}(\alpha)\rangle~~~(\alpha=1,\dots,A),\label{48}
\end{equation}
\emph{viz.},
\begin{equation}
\mid Det (\vec{x}) \rangle=\frac{1}{\sqrt{A!}}\sum_{\mathit{\hat{\mathcal{P}}
\in S_A}} \epsilon_{ \mathcal{P}} \hat{\mathcal{P}}\{\mid \phi_{p_1}(1;\vec{x%
})\rangle \dots \mid \phi_{p_A}(A;\vec{x})\rangle\}.\label{49}
\end{equation}
In turn, such orbitals can be evaluated in a concise analytic form as
initial ones are linear combinations of the HOM orbitals (see Appendix A).

Following (\ref{43}) we arrive to
\[
F_{corr}(\vec{x},\vec{y}) \equiv \langle\Phi_{corr}(-\vec{x})\mid \hat{E}%
_1^\dag(-\vec{y})\hat{E}_1(\vec{y})\mid\Phi_{corr}(\vec{x})\rangle=
\]
\begin{equation}
=\langle Det(-\vec{x})\mid \hat{C}^\dag\hat{E}_1^\dag(-\vec{y})\hat{E}_1(%
\vec{y})\hat{C} \mid Det(\vec{x})\rangle.\label{50}
\end{equation}
Expressions (\ref{29}) and (\ref{32}) with expectations $F(\vec{v},\vec{s})$ and $F(\vec{%
v}~^{\prime},\vec{s}~^{\prime})$, which are determined by eq. (\ref{50}), are
certain base for our calculations.

\subsection{Calculations with the Jastrow-type correlator}
We have seen how expectations (\ref{25}) and (\ref{28}) with respect to the correlated
WF (\ref{38}) can be expressed through the generating function
\begin{equation}
F_{corr}(\vec{x},\vec{y})=\frac 1A\langle Det(-\vec{x})\mid \hat{Q}_{corr}(%
\vec{y})\mid Det(\vec{x})\rangle ,  \label{eq51}
\end{equation}
\begin{equation}
\hat{Q}_{corr}(\vec{y})=\hat{C}^{\dagger }\sum\limits_{\alpha =1}^AE_\alpha
^{\dagger }(-\vec{y})E_\alpha (\vec{y})\hat{C}.  \label{eq52}
\end{equation}
Since we are going to demonstrate (at least, qualitatively) the CMM effects
on the FFs and MDs against the SRCs inclusion (\ref{38}), let us employ, first of
all, the Jastrow ansatz (\ref{39}),
\begin{eqnarray}
\hat{C} &=&\hat{J}=\hat{f}(1,2)\hat{f}(1,3)...\hat{f}(1,A)  \nonumber \\
&&\,\,\,\,\,\,\,\,\,\,\,\,\,\,\,\,\,\,\,\,\,\,\,\times \hat{f}(2,3)...\hat{f}%
(2,A)  \nonumber \\
&&\,\,\,\,\,\,\,\,\,\,\,\,\,\,\,\,\,\,\,\,\,\,\,\,\,\,\,\,\,\,\,\,\,\times
\hat{f}(A-1,A).  \label{eq53}
\end{eqnarray}
Then we have the decomposition
\begin{eqnarray}
\hat{Q}_J(\vec{y})\equiv \hat{J}^{\dagger }\hat{Q}^{[1]}(\vec{y})\hat{J}=%
\hat{Q}^{[1]}(\vec{y})+\hat{Q}^{[2]}(\vec{y})+...\nonumber\\
+\hat{Q}^{[A]}(\vec{y}),
\label{eq54}
\end{eqnarray}
where $\hat{Q}^{[n]}(\vec{y})$ is an $n$-body operator so that
\begin{equation}
\hat{Q}^{[1]}(\vec{y})=\sum\limits_{\alpha =1}^AE_\alpha ^{\dagger }(-\vec{y}%
)E_\alpha (\vec{y}),  \label{eq55}
\end{equation}
\begin{equation}
\hat{Q}^{[2]}(\vec{y})=\sum\limits_{\alpha <\beta }^A\hat{Q}_{\alpha \beta }(%
\vec{y}),  \label{eq56}
\end{equation}
etc.

A systematic way of obtaining separate contributions $\hat{Q}^{[n]}(n\geq 2)$
is prompted by the UCOA (see also \cite{Feld98}, where one can find general analytic expressions
for the corresponding correlated operators). In case of commuting operators $%
\hat{f}(\alpha ,\beta )$ ({\it e.g.}, for the central correlation factors $\hat{f}%
(\alpha ,\beta )=1+h(\mid \widehat{\vec{r}}_\alpha -\widehat{\vec{r}}_\beta
\mid )$ depending only on the distance between particles) one can write (cf.
Appendix A in \cite{ShaWagHu67}),
\begin{equation}
\hat{J}=\exp (\sum\limits_{\alpha <\beta }^A\ln [1+\hat{h}(\alpha ,\beta )]).
\label{eq57}
\end{equation}
After this, applying the UCOM procedure we get
\begin{eqnarray}
\hat{Q}_{\alpha \beta }(\vec{y}) &=&[1+\hat{h}^{\dagger }(\alpha ,\beta )]\{%
\hat{E}_\alpha ^{\dagger }(-\vec{y})\hat{E}_\alpha (\vec{y})\nonumber \\
&&+\hat{E}_\beta
^{\dagger }(-\vec{y})\hat{E}_\beta (\vec{y})\}[1+\hat{h}(\alpha ,\beta )]
\nonumber \\
&&-\hat{E}_\alpha ^{\dagger }(-\vec{y})\hat{E}_\alpha (\vec{y})-\hat{E}%
_\beta ^{\dagger }(-\vec{y})|\hat{E}_\beta (\vec{y})  \label{eq58}
\end{eqnarray}
Along such a guideline we obtain putting in eq. (\ref{eq51}) once $\vec{x}=\vec{v}$
and $\vec{y}=\vec{s}$ by eq. (\ref{31})
\begin{equation}
F_{corr}(\vec{v},\vec{s})=\exp (\frac{q^2r_0^2}4)F_C(\vec{q},\vec{v}),
\label{eq59}
\end{equation}
\begin{equation}
F_C(\vec{q},\vec{v})=\frac 1A\langle Det(-\vec{v})\mid \hat{C}^{\dagger
}\sum\limits_{\alpha =1}^Ae^{i\vec{q}\widehat{\vec{r}}_\alpha }\hat{C}\mid
Det(\vec{v})\rangle ,  \label{eq60}
\end{equation}
and then $\vec{x}=\vec{v}^{\prime }$ and $\vec{y}=\vec{s}^{\prime }$ by eq.
(\ref{34})
\begin{equation}
F_{corr}(\vec{v}^{\prime },\vec{s}^{\prime })=\exp (\frac{z^2p_0^2}4)N_C(%
\vec{z},\vec{v}^{\prime }),  \label{eq61}
\end{equation}
\begin{equation}
N_C(\vec{z},\vec{v}^{\prime })=\frac 1A\langle Det(-\vec{v}^{\prime })\mid
\hat{C}^{\dagger }\sum\limits_{\alpha =1}^Ae^{i\vec{z}\widehat{\vec{p}}%
_\alpha }\hat{C}\mid Det(\vec{v}^{\prime })\rangle ,  \label{eq62}
\end{equation}
When deriving these formulae, we have used the relation,
\begin{equation}
\exp (-\vec{y}^{*}\widehat{\vec{a}}_\alpha \,^{\dagger })\exp (-\vec{y}%
\widehat{\vec{a}}_\alpha )=e^{\frac 12\vec{y}^{*}\vec{y}}\exp [-\vec{y}^{*}%
\widehat{\vec{a}}_\alpha \,^{\dagger }+\vec{y}\widehat{\vec{a}}_\alpha ],
\label{eq63}
\end{equation}
this specific realization of formula (\ref{BakHaus}) for any $c$-vector $\vec{y}$.

Our consideration is simplified if $Det(\vec{x})$ becomes independent of the
vector $\vec{x}$, {\it i.e.},
\begin{equation}
\mid Det(\vec{x})\rangle =\mid Det(0)\rangle =\mid SD\rangle ,  \label{eq64}
\end{equation}
where $\mid SD\rangle $ is an original Slater determinant (see below). Then
\begin{equation}
F_C(\vec{q},\vec{v})=F_C(\vec{q},0)=\frac 1A\langle SD\mid \hat{C}^{\dagger
}\sum\limits_{\alpha =1}^Ae^{i\vec{q}\widehat{\vec{r}}_\alpha }\hat{C}\mid
SD\rangle  \label{eq65}
\end{equation}
and
\begin{equation}
N_C(\vec{z},\vec{v}^{\prime })=N_C(\vec{z},0)=\frac 1A\langle SD\mid \hat{C}%
^{\dagger }\sum\limits_{\alpha =1}^Ae^{i\vec{z}\widehat{\vec{p}}_\alpha }%
\hat{C}\mid SD\rangle .  \label{eq66}
\end{equation}
In accordance with eqs. (\ref{25}) and (\ref{27}) the corresponding FF and MD can be
written as
\begin{equation}
F_{EST}(q)=F_{TB}(q)F_C(\vec{q})  \label{eq67}
\end{equation}
with
\begin{equation}
F_C(\vec{q})=\frac{\langle SD\mid \hat{C}^{\dagger }e^{i\vec{q}\widehat{\vec{%
r}}_1}\hat{C}\mid SD\rangle }{\langle SD\mid \hat{C}^{\dagger }\hat{C}\mid
SD\rangle }  \label{eq68}
\end{equation}
and
\begin{equation}
\eta _{EST}(p)=\frac 1{(2\pi )^3}\int e^{-i\vec{p}\vec{z}}N_{TB}(z)N_C(z)d%
\vec{z}  \label{eq69}
\end{equation}
with
\begin{equation}
N_C(z)=\frac{\langle SD\mid \hat{C}^{\dagger }e^{i\vec{z}\widehat{\vec{p}}_1}%
\hat{C}\mid SD\rangle }{\langle SD\mid \hat{C}^{\dagger }\hat{C}\mid
SD\rangle }  \label{eq70}
\end{equation}
The canonical TB factor
\begin{equation}
F_{TB}(q)=\exp (\frac{q^2r_0^2}{4A})  \label{eq71}
\end{equation}
has appeared in formula (\ref{20}) for the intrinsic operator $\hat{F}_{int}(q)$,
while
\begin{equation}
N_{TB}(z)=\exp (\frac{z^2p_0^2}{4A})  \label{eq72}
\end{equation}
is the own TB factor (see discussion in ref. \cite{ShePaMav06}) for the intrinsic MD.
Respectively, the function $F_C(\vec{q})$ and the Fourier transform
\begin{equation}
\eta _C(p)=\frac 1{(2\pi )^3}\int e^{-i\vec{p}\vec{z}}N_C(z)d\vec{z}
\label{eq73}
\end{equation}
determine the no CM corrected FF and MD with the correlated g.s. (\ref{38})
normalized to unity.

To go on our exploration with Jastrow-type correlations, let us write down
instead of eqs. (\ref{eq60}) and (\ref{eq62}) as in eq. (\ref{eq54}),
\begin{equation}
F_J(\vec{q},\vec{v})=F^{[1]}(\vec{q},\vec{v})+F^{[2]}(\vec{q},\vec{v}%
)+...+F^{[A]}(\vec{q},\vec{v})  \label{eq74}
\end{equation}
and
\begin{equation}
N_J(\vec{z},\vec{v}^{\prime })=N^{[1]}(\vec{z},\vec{v}^{\prime })+N^{[2]}(%
\vec{z},\vec{v}^{\prime })+...+N^{[A]}(\vec{z},\vec{v}^{\prime })
\label{eq75}
\end{equation}
to obtain with the help of the UCOM the following expressions:
\begin{equation}
F^{[1]}(\vec{q},\vec{v})=\frac 1A\langle Det(-\vec{v})\mid
\sum\limits_{\alpha =1}^Ae^{i\vec{q}\widehat{\vec{r}}_\alpha }\mid Det(\vec{v%
})\rangle ,  \label{eq76}
\end{equation}
\begin{eqnarray}
F^{[2]}(\vec{q},\vec{v})=
\frac 1A\langle Det(-\vec{v})\mid
\sum\limits_{\alpha <\beta }^A[\hat{f}^2(\alpha ,\beta )-1]\nonumber\\
\times[e^{i\vec{q}%
\widehat{\vec{r}}_\alpha }+e^{i\vec{q}\widehat{\vec{r}}_\beta }]\mid Det(%
\vec{v})\rangle ,  \label{eq77}
\end{eqnarray}
\[
...\,\,\,\,\,\,\,\,\,\,\,\,\,\,\,\,\,\,\,\,\,\,\,...\,\,\,\,\,\,\,\,\,\,\,\,%
\,\,\,\,\,\,\,\,\,\,\,...\,\,\,\,\,\,\,\,\,\,\,\,\,\,\,\,\,\,\,\,\,\,\,\,\,%
\,\,\,...\,\,\,\,\,\,\,\,\,\,\,\,\,\,\,\,\,\,\,\,\,\,\,\,\,...
\]
and
\begin{equation}
N^{[1]}(\vec{z},\vec{v}^{\prime })=\frac 1A\langle Det(-\vec{v}^{\prime
})\mid \sum\limits_{\alpha =1}^Ae^{i\vec{z}\widehat{\vec{p}}_\alpha }\mid
Det(\vec{v}^{\prime })\rangle ,  \label{eq78}
\end{equation}%
\begin{equation}
\begin{split}
N^{[2]}(\vec{z},\vec{v}^{\prime })&=\frac 1A\langle Det(-\vec{v}^{\prime
})\mid\\
\times&\sum\limits_{\alpha <\beta }^A\{\hat{f}(\alpha ,\beta )[e^{i\vec{z}%
\widehat{\vec{p}}_\alpha }+ e^{i\vec{z}\widehat{\vec{p}}_\beta }]\hat{f}%
(\alpha ,\beta )\\
&-e^{i\vec{z}\widehat{\vec{p}}_\alpha }-e^{i\vec{z}\widehat{%
\vec{p}}_\beta }\}\mid Det(\vec{v}^{\prime })\rangle ,
\end{split}
\label{eq79}
\end{equation}
\[
...\,\,\,\,\,\,\,\,\,\,\,\,\,\,\,\,\,\,\,\,\,\,\,...\,\,\,\,\,\,\,\,\,\,\,\,%
\,\,\,\,\,\,\,\,\,\,\,...\,\,\,\,\,\,\,\,\,\,\,\,\,\,\,\,\,\,\,\,\,\,\,\,\,%
\,\,\,...\,\,\,\,\,\,\,\,\,\,\,\,\,\,\,\,\,\,\,\,\,\,\,\,\,...
\]
for central correlation factor $\hat{f}(\alpha ,\beta )=f(\mid \widehat{\vec{%
r}}_\alpha -\widehat{\vec{r}}_\beta \mid )$ ($\alpha ,\beta =1,...,A$).

\subsection{Application to $^{4}He$}
On the condition (\ref{eq64}) the matrix elements (\ref{eq76})--(\ref{eq79}) are transformed into
the corresponding expectations with respect to the $\mid SD\rangle $. Such a
situation is realized for the pure HOM $(1s)^4$ configuration occupied by
the four nucleons in $^4He$. Indeed, it is the case, where the orbitals
\[
\mid \phi _a(\alpha )\rangle =\mid \varphi _{1s}(\alpha )\rangle \mid \chi
_{\sigma \tau} (\alpha ) \rangle ,
\]
is annuled with the operators $\widehat{\vec{a}}_\alpha $ ($\alpha =1,...,4$%
) so the renormalized orbitals (48)
coincide with the initial $\mid \phi _a(\alpha )\rangle $. Here $\chi
_{\sigma \tau }$ is the spin (isospin) part of the orbital ($\sigma\tau = ++, +-, -+, - - $).
In other words, the corresponding determinant (\ref{49}) does not depend on $\vec{x%
}$, {\it i.e.},
\begin{equation}
\mid Det(\vec{x})\rangle =\mid Det(0)\rangle =\mid (1s)^4\rangle .
\label{eq80}
\end{equation}

Taking into account the definitions (\ref{eq68}) and (\ref{eq70}), the quantities in
question can be represented as the ratios,
\begin{equation}
F_{C}(q)=F_{J}(q)=\frac{A_{J}(q)}{A_{J}(0)}  \label{eq81}
\end{equation}%
and
\begin{equation}
N_{C}(z)=N_{J}(z)=\frac{B_{J}(z)}{B_{J}(0)},  \label{eq82}
\end{equation}%
where
\begin{eqnarray}
A_{J}(q) &=&\langle (1s)^{4}\mid \hat{J}^{\dagger }e^{i\vec{q}\widehat{\vec{r%
}}_{1}}\hat{J}\mid (1s)^{4}\rangle  \nonumber \\
&=&A^{[1]}(q)+A^{[2]}(q)+...+A^{[A]}(q)  \label{eq83}
\end{eqnarray}%
and
\begin{eqnarray}
B_{J}(z) &=&\langle (1s)^{4}\mid \hat{J}^{\dagger }e^{i\vec{z}\widehat{\vec{p%
}}_{1}}\hat{J}\mid (1s)^{4}\rangle  \nonumber \\
&=&B^{[1]}(z)+B^{[2]}(z)+...+B^{[A]}(z)  \label{eq84}
\end{eqnarray}%
so that $B_{J}(0)=A_{J}(0)$.

One should point out that we prefer to deal with finite decompositions (\ref{eq83})
and (\ref{eq84}) retaining
for our approximations only a few first terms of them. Effects of the
neglected terms can be estimated (at least, for $^{4}He$ as in \cite{DalStringBoh82}) by means of
a direct computation without any decomposition (see sec. 5). Of course, the
numerator and denominator in each ratio (\ref{eq81}) and (\ref{eq82}) should be equally
truncated to meet the requirements $F_{J}(0)=1$ and $N_{J}(0)=1$, which
guarantee the correct normalization of DDs and MDs. In the context, we will
recall many works based upon the so-called $\eta $-expansion (see paper \cite{Alviol05}
and refs. therein) of the inverse denominator $A_{J}^{-1}(0)$ in a series.
In our opinion, such a procedure create some problem of convergence even for
finite $A$.

Thus we assume
\begin{equation}
A_{J}(q)=A^{[1]}(q)+A^{[2]}(q),  \label{eq85}
\end{equation}%
\begin{equation}
B_{J}(z)=B^{[1]}(z)+B^{[2]}(z)  \label{eq86}
\end{equation}%
with
\begin{equation}
A^{[1]}(q)=\langle e^{i\vec{q}\widehat{\vec{r}}_{1}}\rangle =\int \varphi
_{1s}^{2}(\vec{r})e^{i\vec{q}\vec{r}}d\vec{r},  \label{eq87}
\end{equation}%
\begin{equation}
B^{[1]}(z)=\langle e^{i\vec{z}\widehat{\vec{p}}_{1}}\rangle =\int \tilde{%
\varphi}_{1s}^{2}(\vec{p})e^{i\vec{z}\vec{p}}d\vec{p},  \label{eq88}
\end{equation}%
\begin{equation}
A^{[2]}(q)=\frac{1}{A}\langle \sum\limits_{\alpha <\beta }^{A}\hat{A}%
_{\alpha \beta }(\vec{q})\rangle =\frac{A-1}{2}\langle \hat{A}_{12}(\vec{q}%
)\rangle  \label{eq89}
\end{equation}%
and
\begin{equation}
B^{[2]}(z)=\frac{1}{A}\langle \sum\limits_{\alpha <\beta }^{A}\hat{B}%
_{\alpha \beta }(z)\rangle =\frac{A-1}{2}\langle \hat{B}_{12}(\vec{z})\rangle.
\label{eq90}
\end{equation}%
Here
\begin{eqnarray}
\hat{A}_{\alpha \beta }(q&&)=\exp [\frac{1}{2}i\vec{q}(\widehat{\vec{r}}%
_{\alpha }+\widehat{\vec{r}}_{\beta })]\nonumber\\
&&\times\{[\hat{f}^{2}(\alpha ,\beta )-1]e^{%
\frac{1}{2}i\vec{q}(\widehat{\vec{r}}_{\alpha }-\widehat{\vec{r}}_{\beta
})}+H.c.\},
\label{eq91}
\end{eqnarray}%
\begin{equation}
\begin{split}
&\hat{B}_{\alpha \beta }(z)=\exp [\frac{1}{2}i\vec{z}(\widehat{\vec{p}}%
_{\alpha }+\widehat{\vec{p}}_{\beta })] \\
&\times\{\hat{f}(\alpha ,\beta )e^{\frac{1}{2%
}i\vec{z}(\widehat{\vec{p}}_{\alpha }-\widehat{\vec{p}}_{\beta })}\hat{f}%
(\alpha ,\beta )-e^{\frac{1}{2}i\vec{z}(\widehat{\vec{p}}_{\alpha }-\widehat{%
\vec{p}}_{\beta })}+H.c.\},
\label{eq92}
\end{split}
\end{equation}%
\[
(\alpha ,\beta =1,...,A)
\]%
since $[\hat{f}(\alpha ,\beta ),\widehat{\vec{r}}_{\alpha }+\widehat{\vec{r}}%
_{\beta }]=[\hat{f}(\alpha ,\beta ),\widehat{\vec{p}}_{\alpha }+\widehat{%
\vec{p}}_{\beta }]=0$, and the symbol $\langle ...\rangle $ is used to
denote the expectation with respect to the determinant $\mid (1s)^{4}\rangle
$ (generally a $\mid SD\rangle $). In eqs. (\ref{eq87})--(\ref{eq88})\, $\varphi _{1s}(\vec{r}%
) $ ($\tilde{\varphi}_{1s}(\vec{p})$) is the 1s orbital in coordinate
(momentum) representation. For convenience, the general HOM orbitals are
given in Appendix A.

Further, calculations by formulae (\ref{B.3})--(\ref{B.8}) with the HOM orbital $%
\varphi _{1s}$ and the correlation factor (\ref{B.11}) are reduced to simple
quadratures. In particular, the approximation (\ref{eq85}) results in the FF,
\begin{equation}
F_J(q)=\frac{A_J(q)}{A_J(0)},  \label{eq93}
\end{equation}
\begin{equation}
A_J(q)=\alpha _1\exp (-\frac{q^2}{4b_1^2})+\alpha _2\exp (-\frac{q^2}{4b_2^2}%
)+\alpha _3\exp (-\frac{q^2}{4b_3^2})  \label{eq94}
\end{equation}
with the coefficients
\[
\alpha _1=1,\,\,\,\,\,\,\alpha _2=-\frac 6{(1+2y)^{3/2}},\,\,\,\,\,\alpha
_3=\frac 3{(1+4y)^{3/2}}
\]
and the falloff parameters
\[
b_1=r_0^{-1}=p_0,\,\,\,b_2=b_1\sqrt{\frac{1+2y}{1+y}},\,\,\,\,b_3=b_1\sqrt{%
\frac{1+4y}{1+2y}}.
\]
\[
b_1<b_2<\,b_3
\]
The DD associated with FF (\ref{eq93}), {\it i.e.}, its Fourier transform, can be
represented as
\begin{equation}
\begin{split}
&\rho _{J}(r)=\frac{\pi ^{-3/2}b_{1}^{3}}{A_{J}(0)}\\
&\times[d_{1}\exp(-b_{1}^{2}r^{2})+d_{2}\exp (-b_{2}^{2}r^{2})+d_{3}\exp (-b_{3}^{2}r^{2})]
\end{split}
\label{eq96}
\end{equation}%
\[
d_{1}=1,\,\,\,\,\,\,d_{2}=-\frac{6}{(1+y)^{3/2}},\,\,\,\,\,d_{3}=\frac{3}{%
(1+2y)^{3/2}}.
\]%

At the same time the approximation (\ref{eq86}) gives rise to the MD (cf. eq. (\ref{eq73})),
\begin{equation}
\begin{split}
&\eta _{J}(p)\equiv \frac{1}{(2\pi )^{3}}\int e^{-i\vec{p}\vec{z}}N_J (z)d%
\vec{z}=\frac{\pi ^{-3/2}b_{1}^{-3}}{A_{J}(0)}\\
&\times[\beta _{1}\exp (-\frac{1}{\gamma _{1}}\frac{p^{2}}{b_{1}^{2}}%
)+\beta _{2}\exp (-\frac{1}{\gamma _{2}}\frac{p^{2}}{b_{1}^{2}})+\beta
_{3}\exp (-\frac{1}{\gamma _{3}}\frac{p^{2}}{b_{1}^{2}})],
\end{split}
\label{eq95}
\end{equation}%
with
\[
\beta _{1}=1,\,\,\,\,\,\,\beta _{2}=-\frac{6}{(1+3y)^{3/2}},\,\,\,\,\,\beta
_{3}=\frac{3}{[(1+4y)(1+2y)]^{3/2}},
\]%
and
\[
\gamma _{1}=1,\,\,\,\,\,\,\gamma _{2}=\frac{1+3y}{1+2y},\,\,\,\,\,\gamma
_{3}=1+2y.
\]%
Henceforth we introduce the dimensionless parameter
\[
y=(\frac{r_{0}}{r_{c}})^{2}
\]

The corresponding CM corrected quantities are determined by
\begin{equation}
F_{J,\emph{EST}}(q)=F_{TB}(q)F_{J}(q),  \label{eq97}
\end{equation}%
\begin{equation}
\rho _{J,\emph{EST}}(r)\equiv \frac{1}{(2\pi )^{3}}\int e^{-i\vec{q}\vec{r}%
}F_{J,\emph{EST}}(q)d\vec{q},  \label{eq99}
\end{equation}%
\begin{equation}
\eta _{J,\emph{EST}}(p)\equiv \frac{1}{(2\pi )^{3}}\int e^{-i\vec{p}\vec{z}%
}N_{TB}(z)N_{J}(z)d\vec{z},  \label{eq98}
\end{equation}%
so
\begin{equation}
\begin{split}
&\rho _{J,\emph{EST}}(r)=\frac{\pi ^{-3/2}b_{1}^{3}}{A_{J}(0)}\\
&\times[\bar{d}_{1}\exp (-\bar{b}_{1}^{2}r^{2})+\bar{d}_{2}\exp (-\bar{b}_{2}^{2}r^{2})+%
\bar{d}_{3}\exp (-\bar{b}_{3}^{2}r^{2})]
\end{split}
\label{eq101}
\end{equation}%
with
\[
\bar{d}_{1}=\left(\frac{\bar{b_1}}{b_1}\right)^{3}\alpha_1,\,\,\,\,\,\,\bar{d}_{2}=\left(\frac{\bar{b_2}}{b_1}\right)%
^{3}\alpha_2,\,\,\,%
\bar{d}_{3}=\left(\frac{\bar{b_3}}{b_1}\right)%
^{3}\alpha_3,
\]%
where
\[
\bar{b}_{1}=\frac{b_1}{\sqrt{1-A^{-1}}},\,\,\,%
\bar{b}_{2}= \frac{b_2}{\sqrt{1-\left( \frac{b_2}{b_1}\right)^2 A^{-1}}},%
\]
\[\bar{b}_{3}= \frac{b_3}{\sqrt{1-\left( \frac{b_3}{b_1}\right)^2 A^{-1}}}%
\]
and
\begin{equation}
\begin{split}
&\eta _{J,\emph{EST}}(p)=\frac{\pi ^{-3/2}b_{1}^{-3}}{A_{J}(0)}\\
&[\bar{\beta}_{1}\exp (-\frac{1}{\bar{%
\gamma}_{1}}\frac{p^{2}}{b_{1}^{2}})+\bar{\beta}_{2}\exp (-\frac{1}{\bar{%
\gamma}_{2}}\frac{p^{2}}{b_{1}^{2}})+\bar{\beta}_{3}\exp (-\frac{1}{\bar{%
\gamma}_{3}}\frac{p^{2}}{b_{1}^{2}})]
\end{split}
\label{eq100}
\end{equation}
with
\[
\bar{\beta}_{1}= \beta_1\left(\frac{\gamma_1}{\bar{\gamma_1}}\right)^{3/2},\,
\bar{\beta}_{2}=\beta_2\left(\frac{\gamma_2}{\bar{\gamma_2}}\right)^{3/2},\,\
\bar{\beta}_{3}= \beta_3\left(\frac{\gamma_3}{\bar{\gamma_3}}\right)^{3/2},
\]
where
\[
\bar{\gamma}_{1}=1-A^{-1},\,\,\,\,\,\,\bar{\gamma}_{2}=\gamma
_{2}-A^{-1},\,\,\,\,\,\bar{\gamma}_{3}=\gamma _{3}-A^{-1}.
\]
For this consideration the mass number $A=4$. But we preserve in these
formulae the $A^{-1}$ -- dependence to indicate a distinct feature of the CMM
correction.

The approach developed here can be useful even if the ansatz (\ref{eq64}) does not
work. In particular, the separate contributions to the matrix element (\ref{eq51}),
that stem from decomposition (\ref{eq54}), can be expressed through certain s.p.
overlap integrals with arbitrary orbitals $\phi _{a}$. For example, we have
for the one-body contribution (\ref{eq76}),
\begin{equation*}
\begin{split}
&F^{[1]}(\vec{q},\vec{v})=\frac{1}{A}\langle \phi _{a_{1}}(1;-\vec{v})\mid
\langle \phi _{a_{2}}(2;-\vec{v})\mid ...\langle \phi _{a_{A}}(A;-\vec{v}%
)\mid \\
&\times \sum\limits_{\alpha =1}^{A}e^{i\vec{q}\widehat{\vec{r}}_{\alpha
}}\Lambda _{A} \mid \phi _{a_{1}}(1;\vec{v})\rangle \mid \phi _{a_{2}}(2;%
\vec{v})\rangle ...\mid \phi _{a_{A}}(A;\vec{v})\rangle
\end{split}
\end{equation*}%
Remind that the antisymmetrizer
\[
\Lambda _{A}=\sum_{\widehat{\mathcal{P}}\in S_{A}}\epsilon _{\mathcal{P}}%
\widehat{\mathcal{P}}
\]%
acts onto the subscripts of orbitals $\phi _{a}(\alpha ;\vec{x})$. It means
that for $(1s)^{4}$ configuration
\begin{eqnarray*}
\begin{split}
&F_{(1s)^{4}}^{[1]}(\vec{q},\vec{v})=\frac{1}{4}\langle \phi _{++}(1;-\vec{%
v})\mid \langle \phi _{+-}(2;-\vec{v})\mid \langle \phi _{-+}(3;-\vec{v}%
)\mid \\
&\times\langle \phi _{--}(4;-\vec{v})\mid
 \sum\limits_{\alpha =1}^{4}e^{i\vec{q}\widehat{\vec{r}}_{\alpha
}}\Lambda _{4} \mid \phi _{++}(1;\vec{v})\rangle \mid \phi _{+-}(2;\vec{v}%
)\rangle \\
&\times\mid \phi _{-+}(3;\vec{v})\rangle \mid \phi _{--}(4;\vec{v})\rangle
\end{split}
\end{eqnarray*}%
whence
\begin{equation}
F_{(1s)^{4}}^{[1]}(\vec{q},\vec{v})=\langle \varphi _{1s}^{\prime }(-\vec{v}%
)\mid e^{i\vec{q}\widehat{\vec{r}}}\mid \varphi _{1s}^{\prime }(\vec{v}%
)\rangle \langle \varphi _{1s}^{\prime }(-\vec{v})\mid \varphi _{1s}^{\prime
}(\vec{v})\rangle ^{3}  \label{eq102}
\end{equation}%
where $\mid \varphi _{1s}^{\prime }(\alpha ;\vec{v})\rangle =\hat{E}_{\alpha
}(\vec{v})\mid \varphi _{1s}(\alpha )\rangle $ ($\alpha =1,2,3,4$) the
renormalized s.p. state and omitting the label $\alpha $ we denote $\mid
\varphi _{1s}^{\prime }(\vec{v})\rangle =E(\vec{v})\mid \varphi _{1s}\rangle
$ (cf. eq. (\ref{48})). Analogously, one can get
\begin{equation}
\begin{split}
&F_{(1s)^{4}}^{[2]}(\vec{q},\vec{v}) =\frac{3}{2}\langle \varphi
_{1s}^{\prime }(1;-\vec{v})\mid \langle \varphi _{1s}^{\prime }(2;-\vec{v}%
)\mid \\
&\times\lbrack \hat{f}^{2}(1,2)-1][e^{i\vec{q}\widehat{\vec{r}}_{1}}+e^{i\vec{%
q}\widehat{\vec{r}}_{2}}]  \\
&\times \mid \varphi _{1s}^{\prime }(1;\vec{v})\rangle \mid \varphi
_{1s}^{\prime }(2;\vec{v})\rangle \langle \varphi _{1s}^{\prime }(-\vec{v}%
)\mid \varphi _{1s}^{\prime }(\vec{v})\rangle ^{2}  \label{eq103}
\end{split}
\end{equation}%
Let us stress once more that if the vector $\mid \varphi _{1s}\rangle $ is a
linear combination of the Cartesian states $\mid n_{x}n_{y}n_{z}\rangle $
the s.p. matrix elements involved are calculated using purely algebraic
means.

In addition, we would like to show some results obtained with the Darmstadt
(D) correlator, which is determined by eqs. (\ref{40})--(\ref{42}). It is the case,
where, {\it e.g.}, instead of the operator $\hat{A}_{12}(\vec{q})$ in expectation
(\ref{eq89}) one should write,
\begin{equation}
\hat{A}_{12}^{D}(\vec{q})=e^{i\vec{q}\widehat{\vec{R}}}\{e^{i\hat{g}(1,2)}e^{%
\frac{1}{2}i\vec{q}\widehat{\vec{r}}}e^{-i\hat{g}(1,2)}-e^{\frac{1}{2}i\vec{q%
}\widehat{\vec{r}}}+H.c.\}  \label{eq104}
\end{equation}%
For brevity, we introduce the CM coordinate $\vec{R}=\frac{1}{2}(\vec{r}_{1}+%
\vec{r}_{2})$ of particles 1 and 2 with their relative coordinate $\vec{r}=%
\vec{r}_{1}-\vec{r}_{2}$ and momentum $\vec{p}=\frac{1}{2}(\vec{p}_{1}-\vec{p%
}_{2})$.

The hermitian generator used in \cite{Feld98} looks as
\begin{equation}
\hat{g}(1,2)=\frac{1}{2}\{\frac{s(\hat{r})}{\hat{r}}\widehat{\vec{r}}%
\widehat{\vec{p}}+\widehat{\vec{p}}\widehat{\vec{r}}\frac{s(\hat{r})}{\hat{r}%
}\}  \label{eq105}
\end{equation}%
One expects the unitary operator $\hat{c}=\exp [-i\hat{g}(1,2)]$ to shift
the relative distance $r$ between the particles {\it via} the position-dependent
displacement $s(r)$. A key point is to find an appropriate function $s(r)$
such that $\hat{c}(1,2)$ could be tractable as a correlator in coordinate
space. In the context, the authors of work \cite{Feld98} have shown that
\begin{equation}
\widehat{\vec{r}}_{g}\equiv \hat{c}(1,2)\widehat{\vec{r}}\hat{c}(1,2)=\frac{%
R_{+}(\hat{r})}{\hat{r}}\widehat{\vec{r}},  \label{eq106}
\end{equation}%
where the shift $R_{+}(r)-r$ characterizes some deviation of the transformed
distance $r_{g}$ from the uncorrelated original $r$.

The relationship (\ref{eq106}) enables us to write
\begin{equation}
\hat{c}^{\dagger }(1,2)e^{\frac{1}{2}i\vec{q}\widehat{\vec{r}}}\hat{c}%
(1,2)=\exp [\frac{1}{2}i\frac{R_{+}(\hat{r})}{\hat{r}}\vec{q}\widehat{\vec{r}%
}]  \label{eq107}
\end{equation}

Substituting (\ref{eq107}) into eq. (\ref{eq104}), we obtain with the $(1s)^{4}$
configuration,
\begin{equation}
A_{D}^{[2]}(q)=\frac{3}{2}\langle \hat{A}_{12}^{D}(\vec{q})\rangle =3\frac{%
\exp [-\frac{1}{8}q^{2}r_{0}^{2}]}{2\sqrt{2}}C(q),  \label{eq108}
\end{equation}%
\[
C(q)=\frac{8\pi }{q}\int\limits_{0}^{\infty }r^{2}dre^{-\frac{1}{2}%
p_{0}^{2}r^{2}}\{\frac{\sin \frac{1}{2}qR_{+}(r)}{R_{+}(r)}-\frac{\sin \frac{%
1}{2}qr}{r}\}
\]

The property $C(0)=0$ provides the required value $F_D(0)=1$ of the
corresponding FF,
\begin{equation}
F_D(q)=\frac{A_D(q)}{A_D(0)}=A^{[1]}(q)+ A_{D}^{[2]}(q)  \label{eq109}
\end{equation}

Furthermore, one can find the relation,
\begin{equation}
R_{+}(r)=r+\Phi (1,2;s\partial _{r})s(r) = r + \int\limits_0^1 du \exp(u s \partial_{r})s(r)
\label{eq110}
\end{equation}%
As anticipated, for a smooth shift function $s(r)$ small compared to $r$
from (\ref{eq110}) it follows (cf. eq. (62) in \cite{Feld98}),
\begin{equation}
R_{+}(r)=r+s(r)+...  \label{eq111}
\end{equation}%
One should note that the authors of \cite{Feld98} not indicating any model for $s(r)$
have preferred to work with the correlation function $R_{+}(r)$ directly.
Our calculation with a parameterized (sophisticated) form for $R_{+}(r)$,
taken from \cite{Feld98}, will be presented somewhere else.

\subsection{Application to $^{16}O$}

For another $j$-closed nucleus $^{16}O$ we will start with the fully
occupied $(1s)^{4}(1p)^{12}$ configuration which is built from the
corresponding HOM orbitals in the $ls$-coupling scheme (see Appendix A).
Now, all we need is to show that the relevant SD (\ref{49}) has the property (\ref{eq64}).
In other words, let us verify the relation
\begin{equation}
\mid Det(\vec{v})\rangle =\hat{E}_{1}(\vec{v})...\hat{E}_{16}(\vec{v})\mid
(1s)^{4}(1p)^{12}\rangle =\mid (1s)^{4}(1p)^{12}\rangle  \label{eq112}
\end{equation}%
for any vector $\vec{v}$.

Indeed, along with the evident equation
\[
\hat{E}(\vec{v})\mid 1s\rangle =e^{\vec{v}\widehat{\vec{a}}}\mid 1s\rangle
=\mid 1s\rangle =\mid 000\rangle \equiv \mid 0\rangle
\]%
we find step by step,
\[
\mid 1p1\rangle =-\frac{1}{\sqrt{2}}\mid 100\rangle -\frac{i}{\sqrt{2}}\mid
010\rangle =(-\frac{1}{\sqrt{2}}\hat{a}_{x}^{\dagger }-\frac{i}{\sqrt{2}}%
\hat{a}_{y}^{\dagger })\mid 0\rangle ,
\]%
\[
\mid 1p0\rangle =\mid 001\rangle =\hat{a}_{z}^{\dagger }\mid 0\rangle ,
\]%
\[
\mid 1p-1\rangle =\frac{1}{\sqrt{2}}\mid 100\rangle -\frac{i}{\sqrt{2}}\mid
010\rangle =(\frac{1}{\sqrt{2}}\hat{a}_{x}^{\dagger }-\frac{i}{\sqrt{2}}\hat{%
a}_{y}^{\dagger })\mid 0\rangle ,
\]%
and
\[
e^{\vec{v}\widehat{\vec{a}}}\mid 1p1\rangle =\mid 1p1\rangle +v_{+1}\mid
1s\rangle ,
\]%
\[
e^{\vec{v}\widehat{\vec{a}}}\mid 1p0\rangle =\mid 1p0\rangle +v_{0}\mid
1s\rangle ,
\]%
\[
e^{\vec{v}\widehat{\vec{a}}}\mid 1p-1\rangle =\mid 1p-1\rangle +v_{-1}\mid
1s\rangle ,
\]%
with the cyclic components
\[
v_{\pm }=\mp \frac{1}{\sqrt{2}}(v_{x}\pm iv_{y}),\,\,v_{0}=v_{z}
\]%
Thus
\begin{equation}
\hat{E}(\vec{v})\mid 1pm\rangle =\mid 1pm\rangle +v_{m}\mid 1s\rangle
,\,\,\,\,\,\,\,(m=1,0,-1)  \label{eq113}
\end{equation}%
Obviously, the second term in the r.h.s. of eq. (\ref{eq113}) does not contribute to
the determinant $\mid D(\vec{v})\rangle $ that immediately gives rise to
(\ref{eq112}).

As before, such an observation essentially simplifies our consideration
since the matrix elements (\ref{eq76})--(\ref{eq79}) and so on are reduced to the
expectations with respect to the customary shell determinant $\mid
(1s)^{4}(1p)^{12}\rangle $. Owing to this, one can again employ formulae (\ref{B.3})--(\ref{B.8}) to get the FFs, DDs, and MDs without any CMM correction,%
\begin{equation}
F_{J}(q)=\frac{A_{J}(q)}{A_{J}(0)},\,\,\,\,\,\ 
 \label{eq114}
\end{equation}%
\begin{equation}
\begin{split}
A_{J}(q)=&\alpha _{1}(q)\exp \left(-\frac{q^{2}}{4b_{1}^{2}}\right)\\
&+\alpha _{2}(q)\exp \left(-%
\frac{q^{2}}{4b_{2}^{2}}\right)+\alpha _{3}(q)\exp \left(-\frac{q^{2}}{4b_{3}^{2}}\right),
\end{split}
\label{eq115}
\end{equation}%

\begin{equation}
\begin{split}
\rho_{J}(r)=\frac{\pi^{-3/2}b_1^{3}}{A_{J}(0)}&
[d_{1}(r)\exp (-b_1^{2}r^{2})\\
&+d_{2}(r)\exp(-b_2^{2}r^{2})-d_{3}(r)\exp(-b_2^{2}r^{2})],
\end{split}
\label{eq131}
\end{equation}

\begin{equation}
\begin{split}
\eta _{J}(p)=&\frac{\pi^{-3/2}b_1^{-3}}{A_{J}(0)}[\beta _{1}(p)\exp (-\frac{1}{\gamma _{1}}\frac{p^{2}}{%
b_{1}^{2}})+\\
&\beta _{2}(p)\exp (-\frac{1}{\gamma _{2}}\frac{p^{2}}{b_{1}^{2}}%
)+\beta _{3}(p)\exp (-\frac{1}{\gamma _{3}}\frac{p^{2}}{b_{1}^{2}})]
\end{split}
\label{eq116}
\end{equation}%
{\it vs.} the CMM corrected ones,%
\begin{equation}
F_{J,\emph{EST}}(q)=F_{TB}(q)F_{J}(q),  \label{eq117}
\end{equation}
\begin{equation}
\begin{split}
\rho_{J,EST}&(r)=\frac{\pi^{-3/2}\bar{b_1}^{3}}{A_{J}(0)}[\bar{d_{1}}(r)\exp (-\bar{b_1}^{2}r^{2})\\
&+\bar{d_{2}}(r)\exp(-\bar{b_2}^{2}r^{2})-\bar{d_{3}}(r)\exp(-\bar{b_2}^{2}r^{2})],
\end{split}
 \label{eq133}
\end{equation}
\begin{equation}
\begin{split}
\eta_{J,EST}&(p)=\frac{\pi^{-3/2}b_1^{-3}}{A_{J}(0)}[\bar{\beta_{1}}(p)\exp (-\frac{p^{2}}%
{b_1^{2}\bar{\gamma_1}})\\
&+\bar{\beta_{2}}(p)\exp(-\frac{p^{2}}{b_1^{2}\bar{\gamma_2}})%
+\bar{\beta_{3}}(p)\exp(-\frac{p^{2}}{b_1^{2}\bar{\gamma_3}})].
\end{split}
\label{eq134}
\end{equation}%
Of course, here we have the relevant TB factor,
\begin{equation}
F_{TB}(q)=\exp (\frac{q^{2}r_{0}^{2}}{64}).\,\,\,\,\,
 \label{eq119}
\end{equation}%
Analytic (in general, cumbersome) expressions for the polynomials $\alpha_{i}(q)$, %
$d_i(r)$, $\bar{d_i}(r)$,  $\beta _{i}(p)$ and  $\bar{\beta_i}(p)$ ($i=1,2,3$)
can be obtained using formulae of Appendix B that results in (by taking, respectively, $x=q/b_1$ and $z=p/b_1$)
\[
\alpha _{1}(q)=1-\frac{x^2}{8},\,\,\,
\alpha _{2}(q)=2\frac{\pi _{11}(y)+\pi _{12}(y)x^2+\pi _{13}(y)x^4}{\left( 1+2y\right) ^{3/2}},\,\,\,
\]%
\[
\alpha _{3}(q)=-\frac{\pi _{21}(y)+\pi _{22}(y)x^2+\pi _{23}(y)x^4}{\left( 1+4y\right) ^{3/2}},
\]
where
\[
\pi _{11}(y)=-1-\frac{50 + 116 y + 77 y^{2}}{4 (1 + 2 y)^{2}},\,\,\,
\]

\[
\pi _{12}(y)=\frac{13 + 25 y + 22 y^{2} + 18 y^{3}}{8 (1 + 2 y)^{3}},\,\,\,
\]

\[
\pi _{13}(y)=\frac{y^{2} (-3 - 2 y + 5 y^{2})}{16 (1 + 2 y)^{4}},
\]%

\[
\pi _{2i}(y)=\left. \pi _{1i}(y)\right\vert _{y\rightarrow 2y} (i=1,2,3)
\]
so
\[
d_{1}(r)=1-\frac{b_1^2 r^2}{2} \left(\frac{3}{2}-b_1^2 r ^2\right) ,
\]

\begin{equation}
\begin{split}
d_{2}&(r)=\frac{2}{(1+y)^{3/2}}(\pi%
_{11}(y)+2\pi _{12}(y) \left(3-2b_2^2 r ^2\right)\left(\frac{b_2}{b_1}\right)^2\\
&+4 \pi _{13}(y)%
\left(15-20 b_2^2 r%
^2+4 b_2^{4} r ^4 \right)\left(\frac{b_2}{b_1}\right)^4) \nonumber\ \ ,%
\end{split}
\end{equation}

\begin{equation}
\begin{split}
&d_{3}(r)=\frac{1}{(1+2y)^{3/2}}(\pi_{21}(y)+2\pi _{22}(y) \left(3-2b_3^2 r^2\right)\left(\frac{b_3}{b_1}\right)^2\\
&+4 \pi _{23}(y)%
    \left(15- 20 b_3^2 r%
     ^2+4b_3^{4} r ^4 \right)\left(\frac{b_3}{b_1}\right)^4)\ \ .%
\nonumber
\end{split}
\end{equation}
At the same time we find for the MD,
\[
\beta_{1}(p)=\frac{1}{4}+\frac{z^2}{2},%
\]

\begin{equation}
\begin{split}
&\beta_{2}(p)=\frac{1}{(1+3 y)^{3/2}}(\eta _{11}(y)+4 \left(\frac{3}{2}-\frac{z^2}{\gamma _2}\right)\left(\eta_{12}(y)/\gamma_2\right)\\
&+8\left(15-\frac{10 z^2}{\gamma%
   _2}+\frac{2 z^4}{\gamma%
   _2^{2}}\right)\left(\eta _{13}(y)/\gamma_2^2\right)),
   \nonumber
\end{split}
\end{equation}

\[
\beta_{3}(p)=\frac{\eta _{21}(y)+4%
   \left(\frac{3}{2}-\frac{z^2}{\gamma _3}\right)\left(\eta _{22}(y)/\gamma_3\right)}{(1+4 y)^{3/2}(1+2 y)^{3/2}}.
\]
where%
\[
\eta _{11}(y)=-\frac{3}{2}\ \ \frac{31 y^2+44 y+18}{(1+2 y)^2},\,\,\,\]

\[
\eta _{12}(y)=\frac{13+69 y+92 y^2+42 y^3}{4 (1+2 y)^3} ,
\]

\[
\eta _{13}(y)=\frac{3}{8}\ \ \frac{ y^2 \left(1+4y+3 y^2\right)}{ (1+2 y)^4},\,\,\,
\]

\[\eta _{21}(y)=\frac{3}{2}\ \ \frac{ 9+44y+62 y^2}{ (1+4 y)^2} ,\,\
\]
\[\eta _{22}(y)=-\frac{ 13+34 y}{8 (1+4 y)}.
\]
The cutoffs $b_{1}$, $b_{2}$, $b_{3}$ , $\gamma _{1}$, $\gamma _{2}$ and $\gamma _{3}$ are
determined as in eq.(\ref{eq94}) and eq.(\ref{eq95}).

The analytic expressions for the polynomials $\bar{d}_i(r)$ and $\bar{\beta}_i(p)$ (i=1,2,3) are obtained by following the recipes:
\[
\bar{d}_1(r)=d_1(r)\vert _{b_1 \rightarrow \bar{b}_1},
\]
\[
\bar{d}_2(r)=\left[1-\frac{1}{A}\, \frac{1+2y}{1+y}\right]^{-3/2} d_2(r)\vert _{b_{1,2} \rightarrow \bar{b}_{1,2}},\,\,
\]
\[
\bar{d}_3(r)=\left[1-\frac{1}{A}\, \frac{1+4y}{1+2y}\right]^{-3/2} d_3(r)\vert _{b_{1,3} \rightarrow \bar{b}_{1,3}},
\]
\[
\bar{\beta}_1(p)=\beta_1(p) \vert _{\gamma_1 \rightarrow \bar{\gamma}_1},
\]
\[
\bar{\beta}_2(p)=\bar{\gamma}_2^{-3/2} \,\beta_2(p) \vert _{\gamma_{1,2} \rightarrow \bar{\gamma}_{1,2}},\,
\bar{\beta}_3(p)=\bar{\gamma}_3^{-3/2} \,\beta_3(p)\vert _{\gamma_{1,3} \rightarrow \bar{\gamma}_{1,3}},
\]
where $\bar{b}_1$, $\bar{b}_2$, $\bar{b}_3$, $\bar{\gamma}_1$, $\bar{\gamma}_2$ and $\bar{\gamma}_3$ are determined in the same way as in the case of $^4 He$.

\section{Results and discussion}

The analytic expressions derived in sect. 4 for density and momentum distributions and their Fourier transforms are sufficiently general to be applied in different translationally invariant treatments with the SRCs included. Our calculations carried out by formulae (\ref{eq93})-(\ref{eq100}) for the $^4 He$ nucleus and by formulae (\ref{eq114})-(\ref{eq119}) for $ ^{16} O$ nucleus are displayed in figs. $1-5$ together with available data. In these figures we distinguish two cases in which
along with the model Jastrow correlations the CMM correction is either included or not.

In order to calculate the charge FFs we have used the relation
\begin{equation}
F_{CH}(q)= F_{TB}(q)F_{DF}(q) F_{proton}(q) F_{int}(q),
\label{FCH}
\end{equation}
 where $F_{DF}(q)=1-q^2/2m^2$ is the Darwin-Foldy correction and $F_{proton}(q)$ is the finite proton size factor with the parametrization from \cite{ChaSau}.

\begin{figure*}
\begin{center}
  \begin{tabular}{cc}
    \resizebox{0.5\textwidth}{!}{\includegraphics{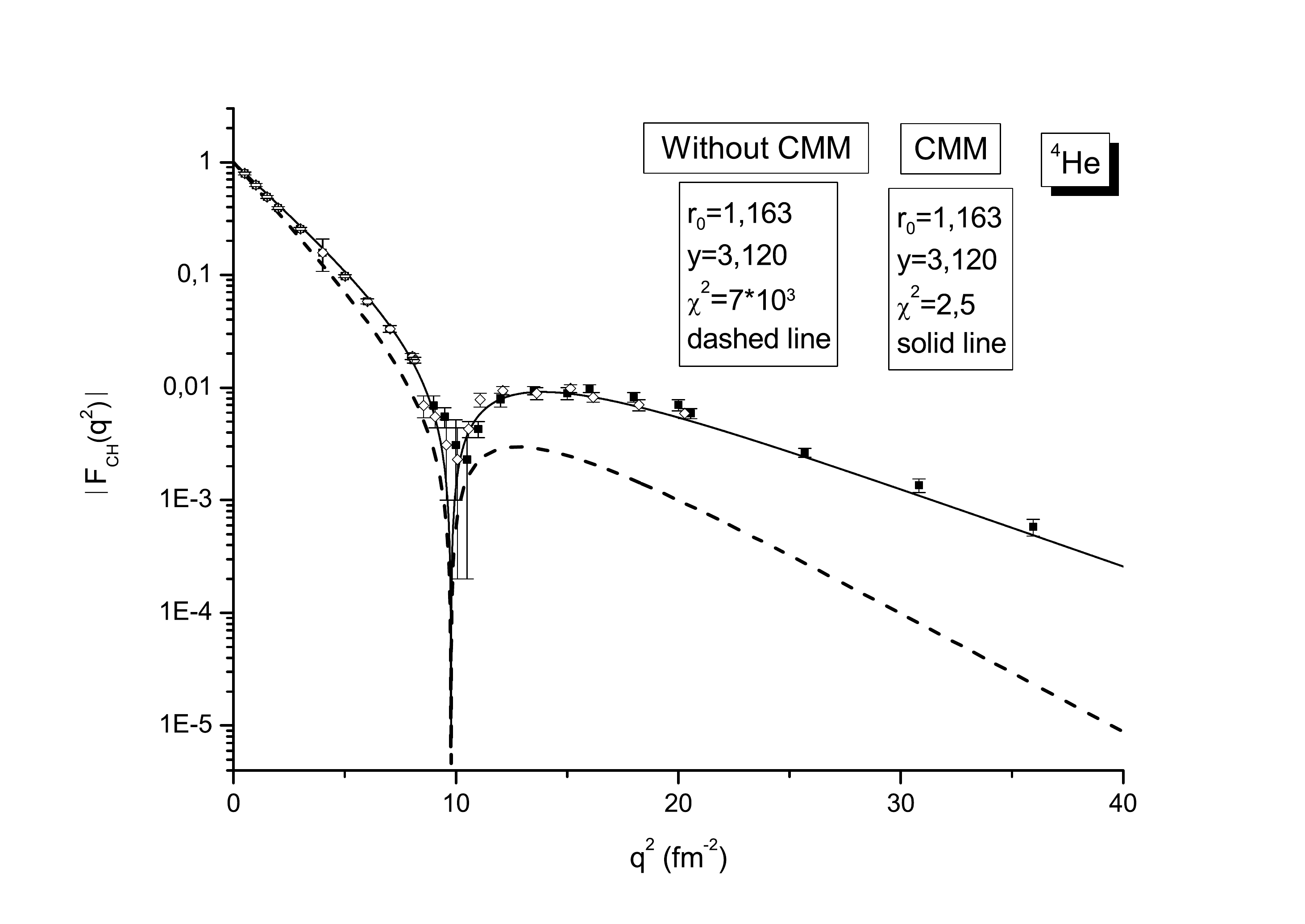}}
  &
    \resizebox{0.5\textwidth}{!}{\includegraphics{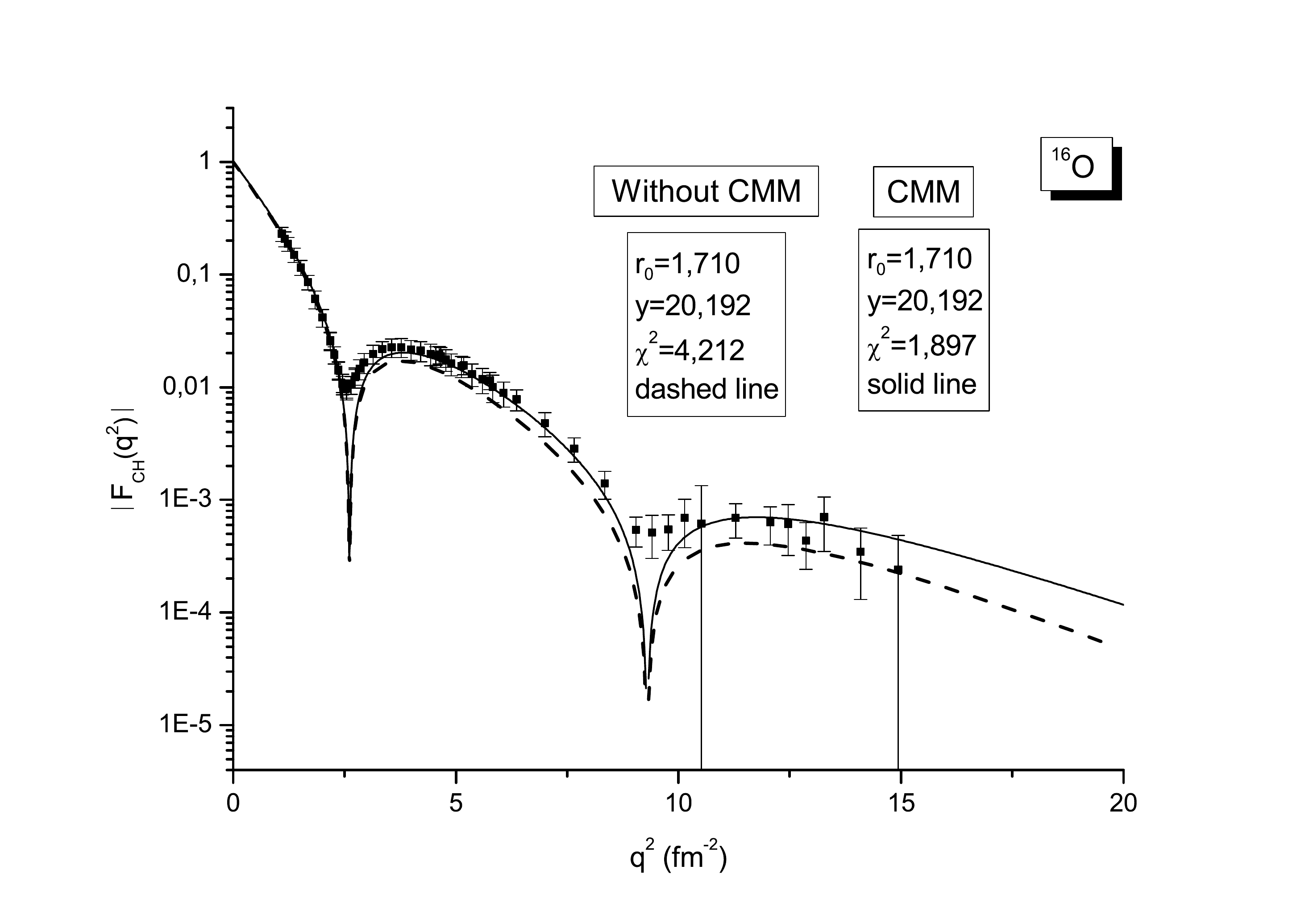}}
  \end{tabular}
\end{center}
\caption{The charge form factor of the nuclei $^4 He$ (on the left) and $^{16} O$ (on the right) : calculated with the Jastrow WF  using the EST prescription (solid curves) and without the CMM correction (dashed curves); experimental points from \cite{FroschArn} and \cite{Sick}, respectively. Other clarifications are given in the text.}
\label{fig:charge FF}
\end{figure*}

 The parameters $r_0$ and $r_c$ (or, equivalently, $y=(r_0/r_c)^2 $) have been extracted from the data in fig. 1 for each nucleus {\it via} a least squares fit to the experimental $F_{CH}(q)$ : their best-fit values are $r_0 =1.163\,fm $ and $y=3.120$ $(r_c=0.658\,fm)$ for $^4 He$ and $r_0 =1.171\,fm$ and $y=20.192$ $(r_c=0.261\,fm)$ for $^{16} O$. Being fixed in such a way, they remain unchanged for subsequent calculations. Along with the best-fit solid curves we have drawn the corresponding dashed curves to demonstrate the CMM influence (sometimes considerable) on the distributions in question.
\begin{figure*}
\begin{center}
  \begin{tabular}{cc}
    \resizebox{0.5\textwidth}{!}{\includegraphics{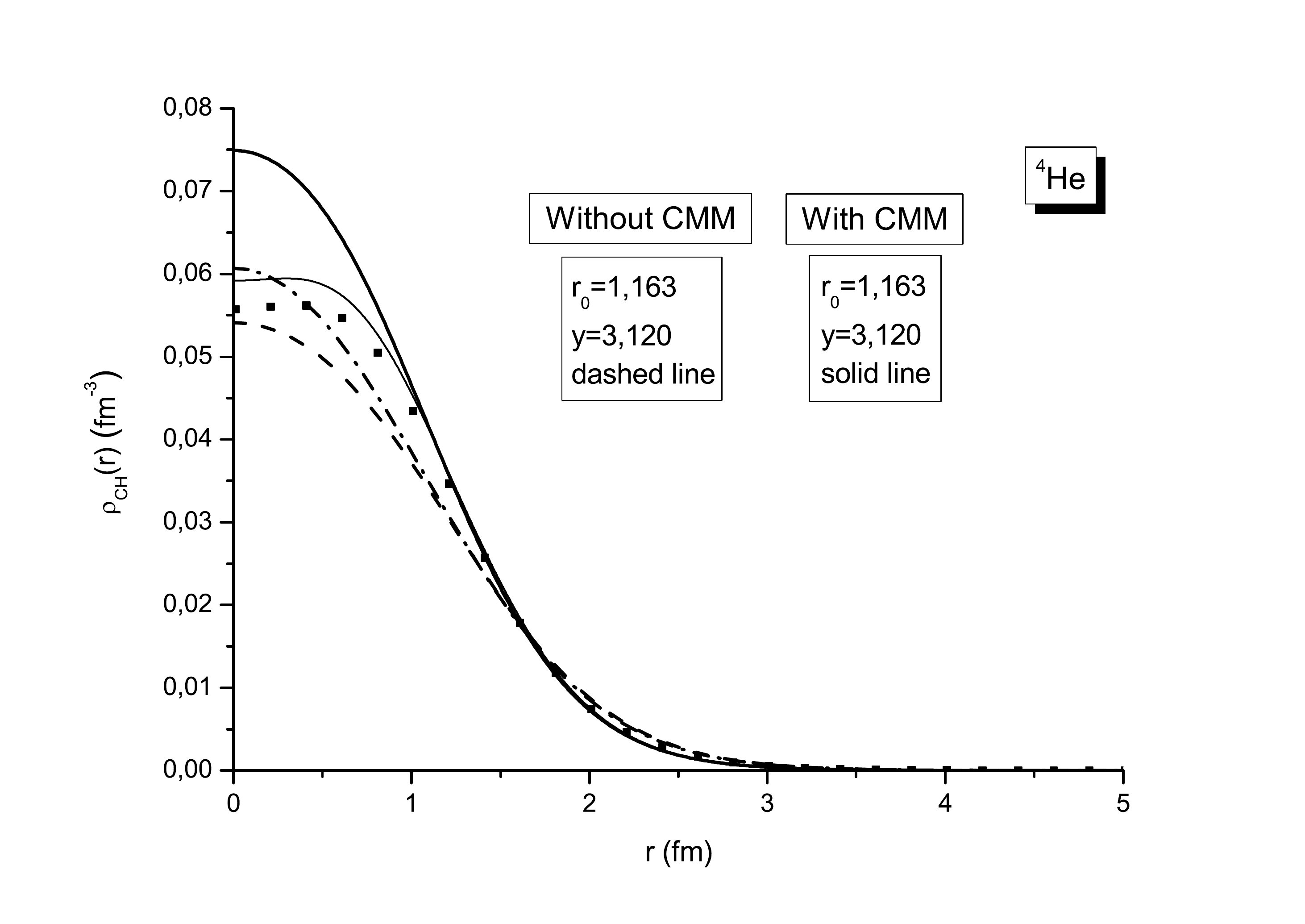}}
  &
    \resizebox{0.5\textwidth}{!}{\includegraphics{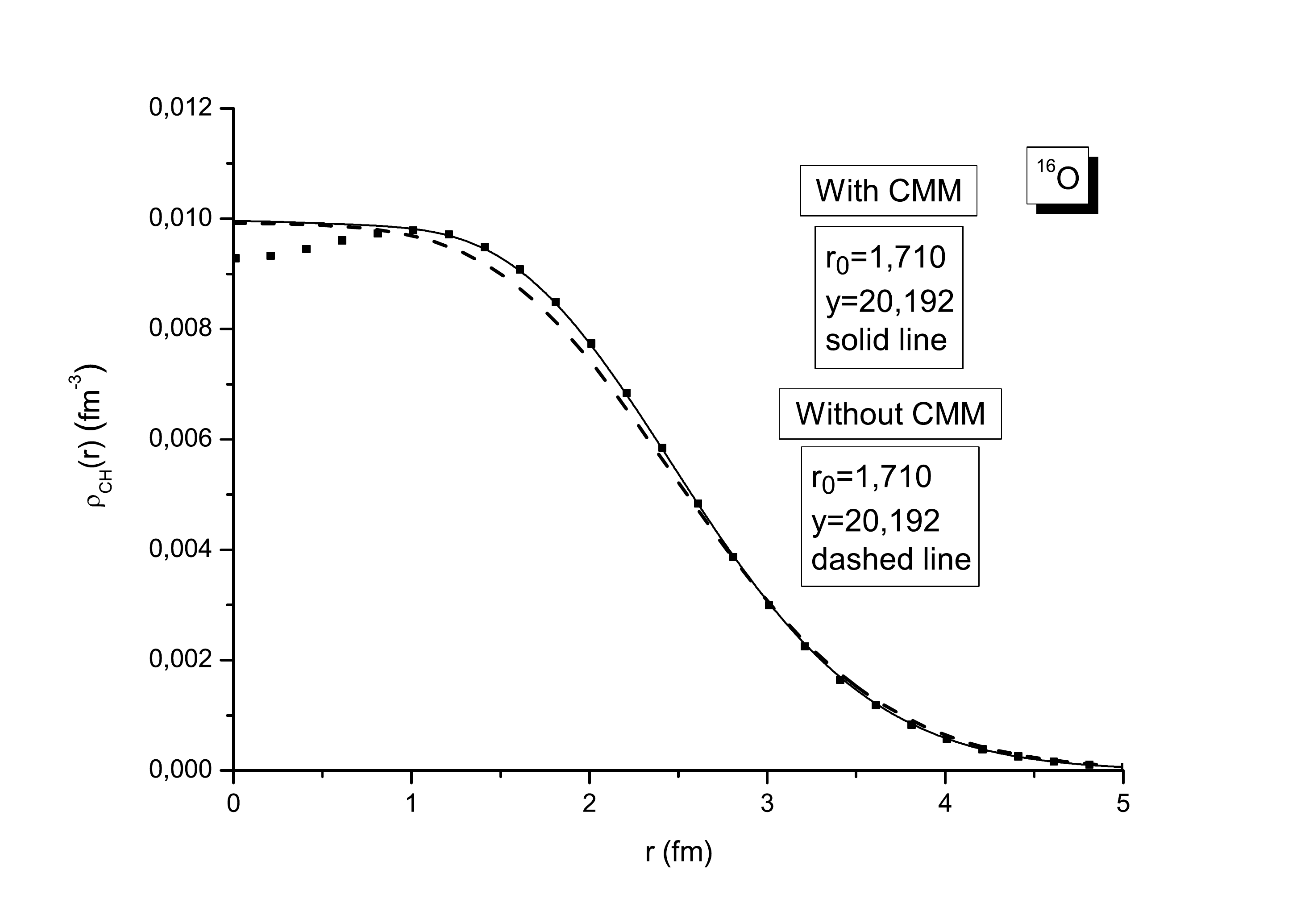}}
  \end{tabular}
\end{center}
\caption{The charge density of the nuclei $^4 He$ and $^{16} O$: calculated with the Jastrow WF  using the EST prescription (solid curves) and without the CMM correction (dashed curves).In addition, the thick solid and dash-dotted curves show our exact (numerical) calculation for $^4 He$, respectively, with the EST prescription and without it; experimental points from \cite{deVries}. Other clarifications are given in the text.}
\label{fig:charge DD}
\end{figure*}
 As seen in fig. 1, the CMM-corrected calculations reproduce the observed q-dependencies of the FFs, {\it viz.}, the envelopes of diffraction maxima and the positions of diffraction minima.

\begin{figure*}
\begin{center}
  \begin{tabular}{cc}
    \resizebox{0.5\textwidth}{!}{\includegraphics{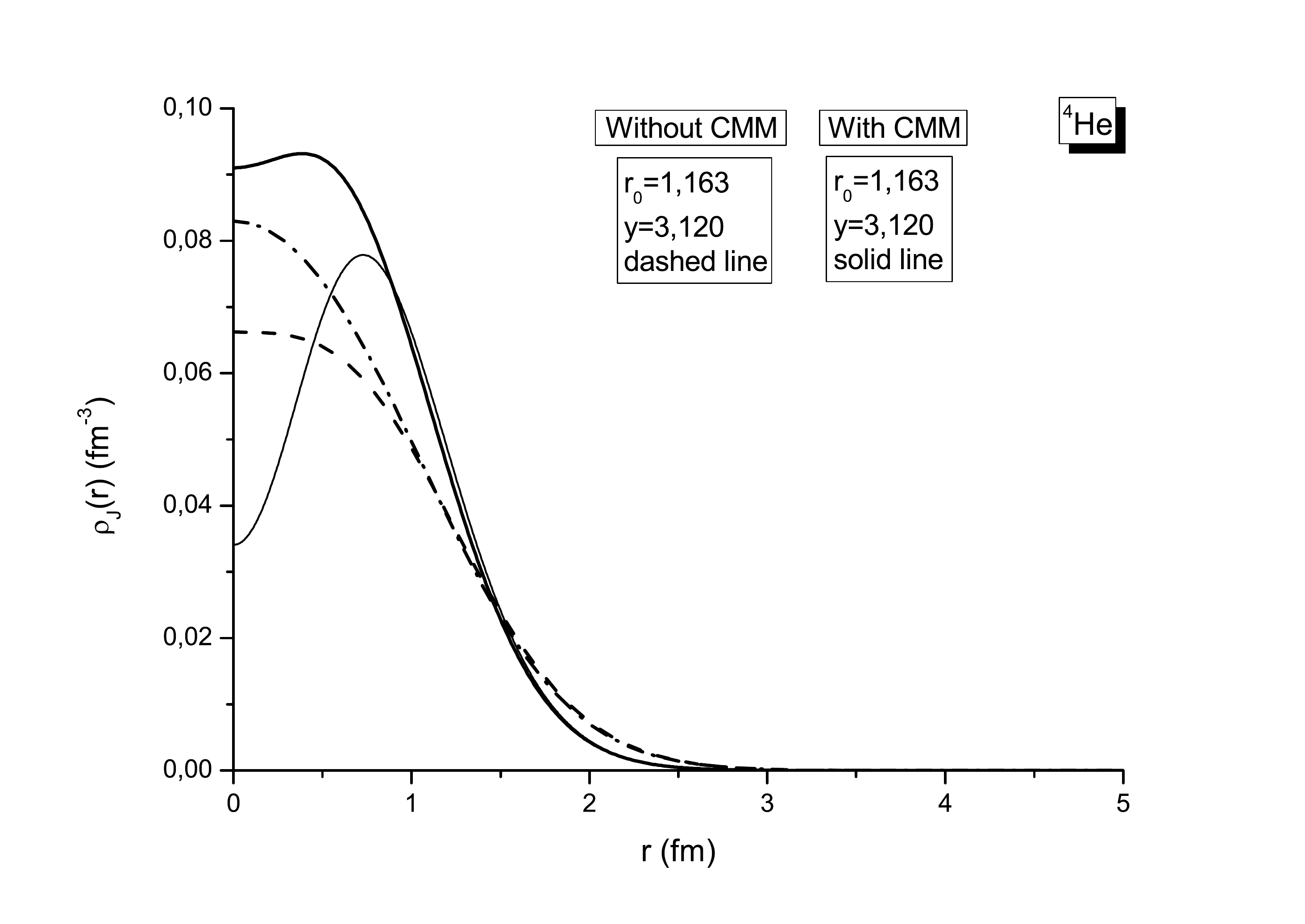}}
  &
    \resizebox{0.5\textwidth}{!}{\includegraphics{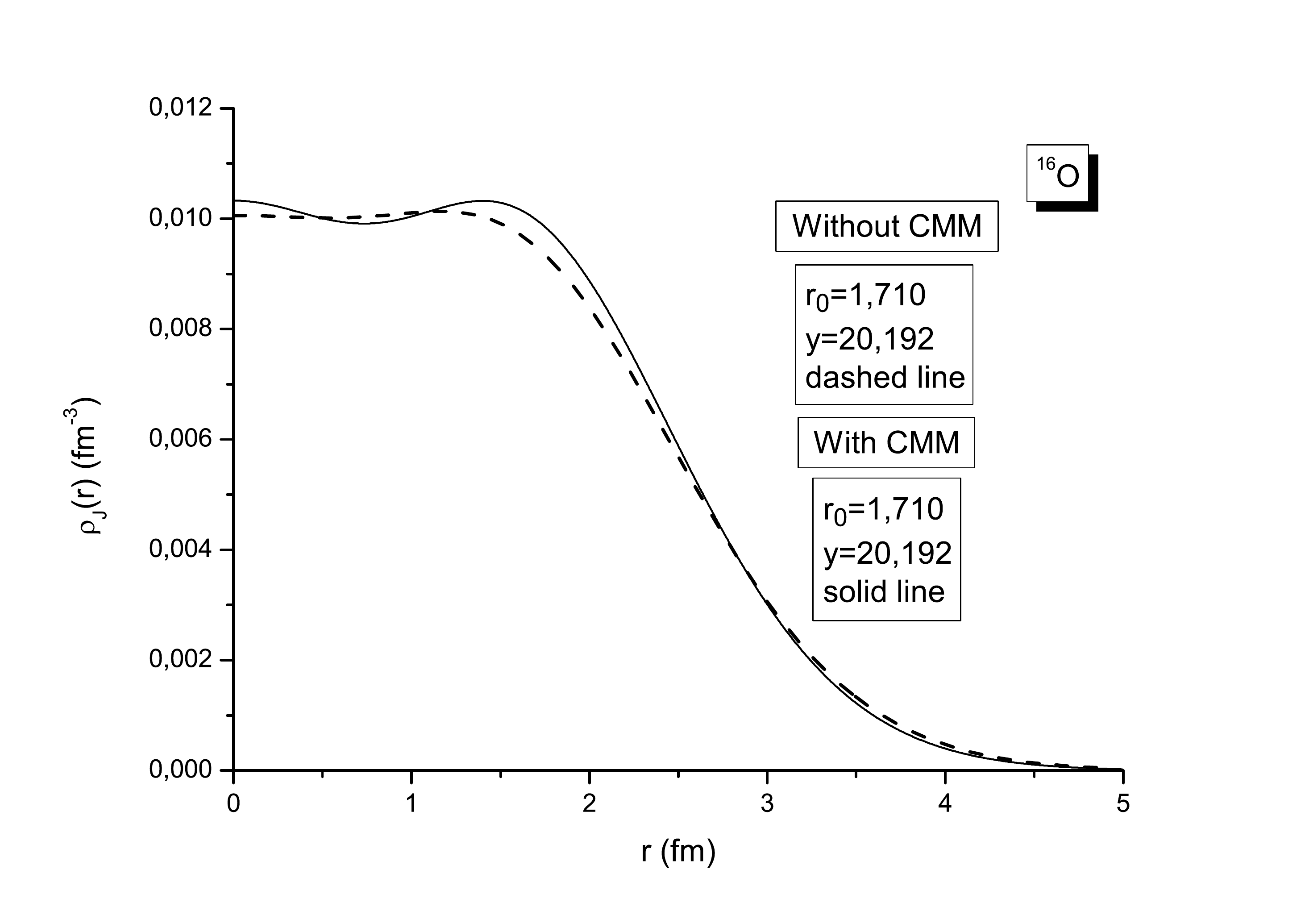}}
  \end{tabular}
\end{center}
\caption{The point-proton density of the nuclei $^4 He$ and $^{16} O$ : calculated by formulae (\ref{eq96}) (dashed curve) and (\ref{eq101}) (solid curve) on the left and by formulae (\ref{eq131}) (dashed curve) and (\ref{eq133}) (solid curve) on the right. Distinctions between other curves are the same as in fig.2. The normalization is $\int\rho_{J}(r) d\vec{r}=1$.}
\label{fig:OBDD}
\end{figure*}

 In order to evaluate validity of the approximation given by eqs. (\ref{eq85})-(\ref{eq86}) we have calculated quantities $A_{J}(q)$ and $B_{J}(z)$ without any truncation of decompositions (\ref{eq83}) and (\ref{eq84}). Comparison between the corresponding curves shows that some qualitative changes of the $r-$ and $p-$ dependencies $\rho_{J,EST}(r)$ and $\eta_{J,EST}(p)$, which are determined, respectively, by (\ref{eq101}) and (\ref{eq100}), can be by-products of the approximation. In fact, considerable dips in the solid curves on the left panels of figs. 3 and 5 do not appear for exact calculations. At the point, one should note that the additional depression of $\rho_{J,EST}(r)$ with respect to $\rho_{J}(r)$  at a moderate $y-$ value ({\it cf}. the solid and dashed curves in fig.3 for the alpha-particle in the range $0\leq r\le 1$) is obscured in the charge density. The latter, being defined as the Fourier transform of the charge FF by formula (\ref{FCH}), is calculated {\it via} the convolution of $\rho_{J,EST}(r)$ with a smoothed charge distribution in the proton. Moreover, it turns out that even with the lack (at smaller $y-$values) of the necessary property of $\rho_{J,EST}(r)$ to be positively definite the convolution results in a distribution $\rho_{CH}(r)$ which has much in common with that shown by the solid curve in the left panel of fig.2. Perhaps, in spite of similar observations many authors (see, {\it e.g.}, Table I in \cite{Massen99} with the parameters $b=r_{0}=1.1732 fm$ and $\beta=r_{c}^{-2}=2.3127 fm^{-2}$ for $^4 He$ that is equivalent to y=3.183) show only the charge densities of nuclei. Further, the exact distribution $\rho_{J,EST}^{exact}(r)$ (the thick solid curve in fig. 3) has a plateau in the vicinity of $r=0$ with a shallow dip. When increasing the $y-$values the $\rho_{J}(r)$ dependencies (both exact and approximate) become smoothly varying functions of the nucleon coordinate $r$.

\begin{figure*}
\begin{center}
   \begin{tabular}{cc}
    \resizebox{0.47\textwidth}{!}{\includegraphics{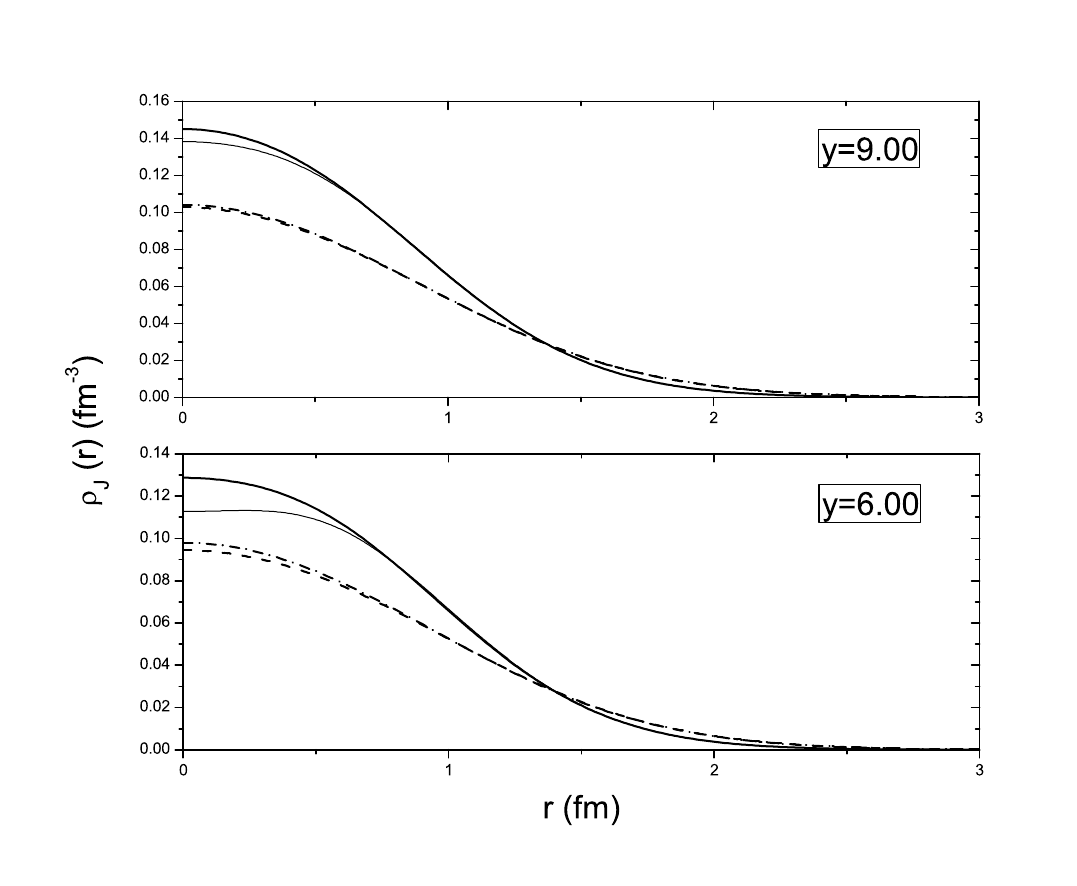}}
  &
    \resizebox{0.47\textwidth}{!}{\includegraphics{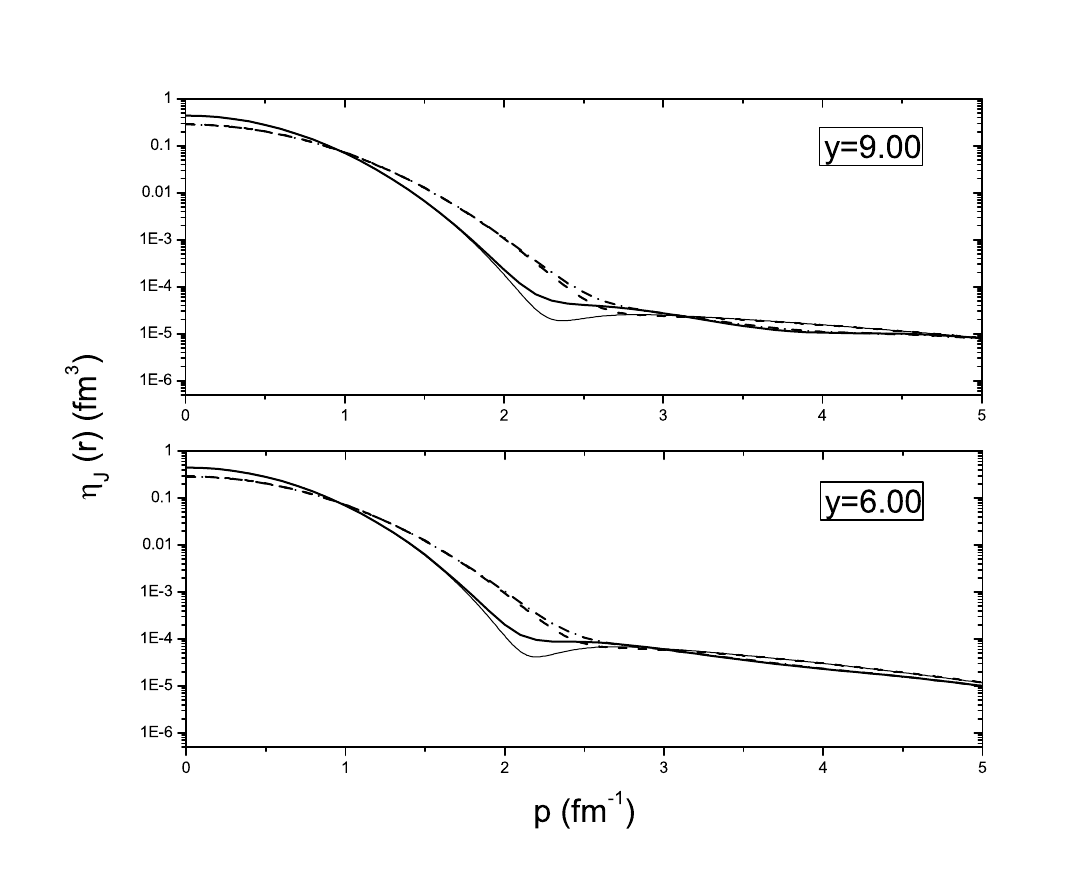}}
   \end{tabular}
\end{center}
\caption{The one-body density (on the left) and momentum distribution (on the right) of the alpha particle at different y-values and fixed $r_0 = 1.163 fm $. As in fig. 3, curves on the left panel calculated by formulae (\ref{eq96}) and (\ref{eq101}) whereas the right panel demonstrates the dependence
$\eta _{J,\emph{EST}}(p)$ (eq. (\ref{eq100}): solid curves) {\it vs.} $\eta _{J}(p)$ (eq. (\ref{eq95}): dashed curves). Distinctions between the thick solid and dash-dotted curves are the same as in fig.3.
The normalization is $\int\eta_{J}(p) d\vec{p}=1$.}
\label{fig:y_dependence}
\end{figure*}

In addition, as seen from figs. 3 and 4, the CMM correction diminishes the expected depression of the intrinsic DD $\rho_{J}(r)$ relative to $\rho_{HOM}(r)=\lim_{y\to\infty} \rho_{J}(r)$ in its central region, {\it i.e.}, increases the probability to find a nucleon in the $^4 He$ interior. From the physical viewpoint such an extra increase is not something exclusive since the TI restoration means the introduction of nucleon-nucleon correlations as a whole (including the short-range ones too).

\begin{figure*}
\begin{center}
   \begin{tabular}{cc}
    \resizebox{0.5\textwidth}{!}{\includegraphics{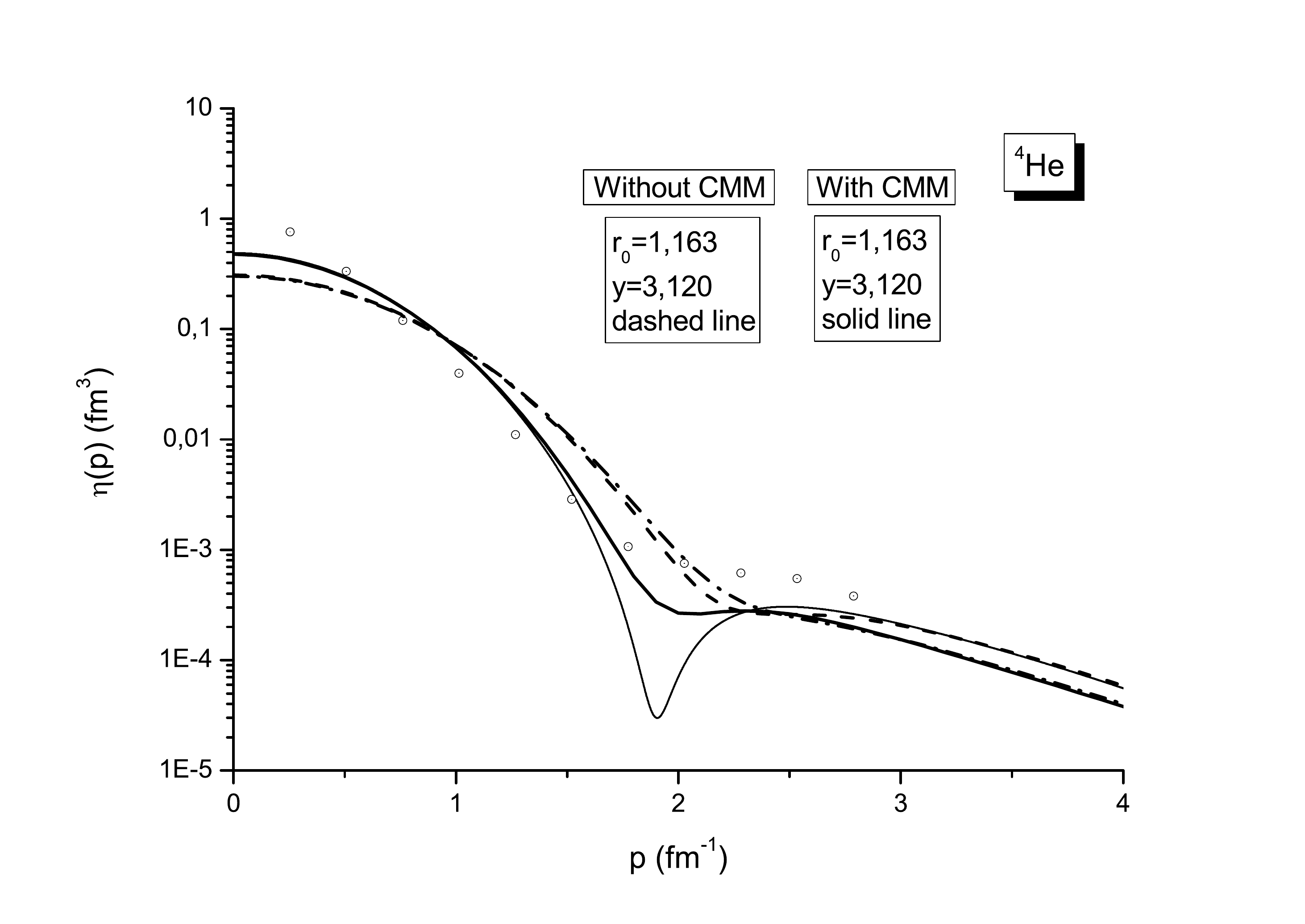}}
  &
    \resizebox{0.5\textwidth}{!}{\includegraphics{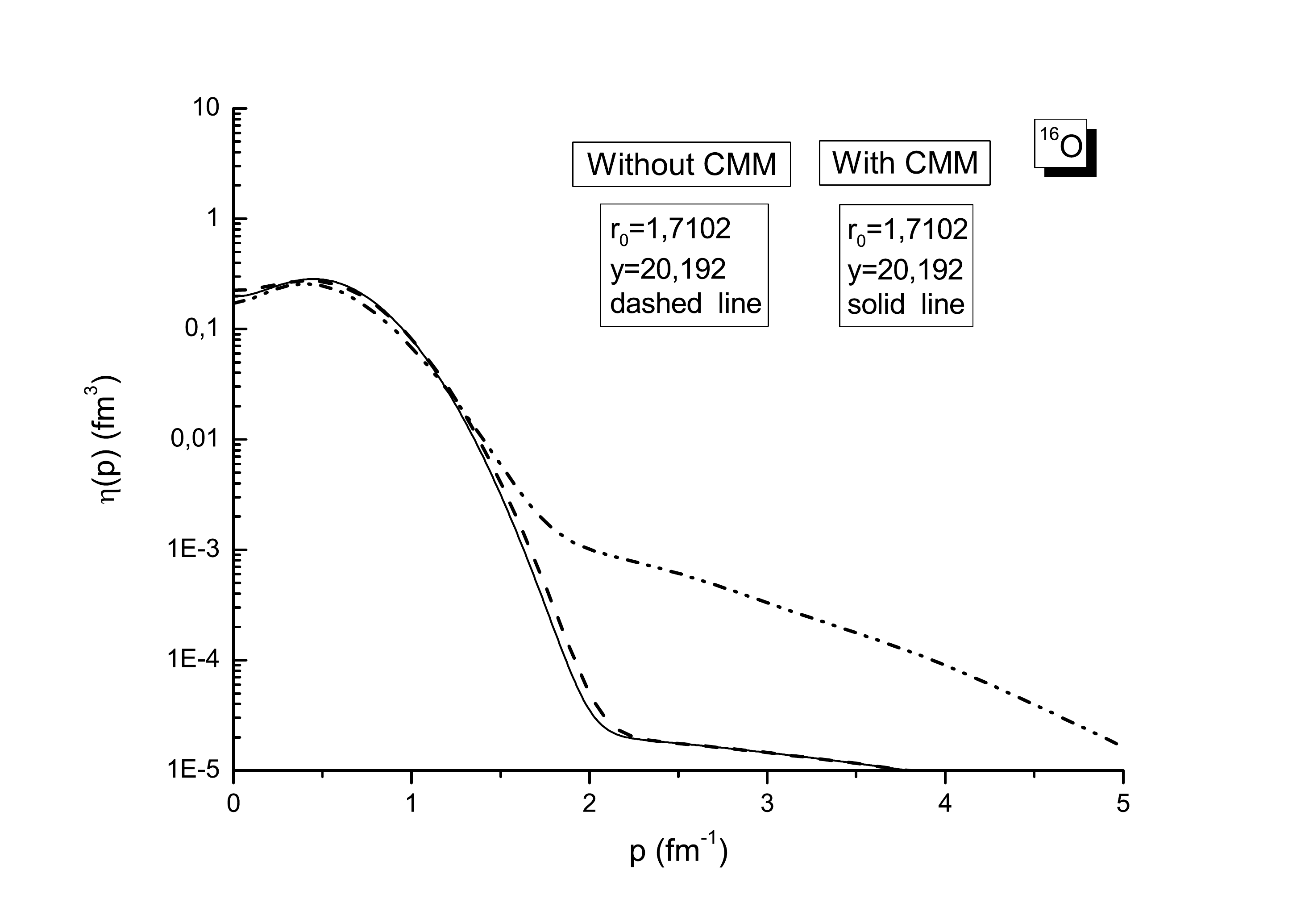}}
   \end{tabular}
\end{center}
\caption{The momentum distributions of the nuclei $^4 He$ and $^{16} O$. Together with our calculations for the best DDs we have depicted the results from \cite{ATMS} (circles on the left) and \cite{Argonne} (the dash-double-dotted curve on the right). Difference between the thick solid and dash-dotted curves explained in fig.3.}
\label{fig:MDs}
\end{figure*}

Going on our discussion of the interplay between the CM fixation and the phenomenological introduction of $N-N$ repulsion in the nuclear wave function, we will note a simultaneous shrinking of the OBDD and OBMD (cf. the thick solid curves {\it vs} dash-dotted ones in figs. 3, 4 and 5). Following \cite{ShePaMav06} the term 'shrinking' implies that the EST prescription gives rise to increasing each of these densities in their central regions (respectively, $0\leq r\le r_0$ and $0\leq p\le p_0=r_{0} ^{-1}$) compared to the nTI quantities. But unlike refs. \cite{ShePaMav06}  and \cite{SheGrig07}, where the effect has been confirmed within the HOM and its modification \cite{RKB} , the present observation is related to the exact numerical results obtained beyond such simple models. In the context, note the relations under the strong inequality $r_c \ll r_0$ with
$$\rho^{EST}_{HOM}(r)= \lim_{y\to\infty}\rho_{J, EST}(r) = \bar{r}^{-3}_{0}\,\pi^{-3/2} \exp(-r^2/\bar{r}^{2}_0) $$
{\it vs.}
$$\rho_{HOM}(r) = r^{-3}_{0}\,\pi^{-3/2}\exp(-r^2/r^{2}_0).$$
and
$$\eta^{EST}_{HOM}(p)= \lim_{y\to\infty}\eta_{J, EST}(p) = \bar{r}^{3}_{0}\,\pi^{-3/2} \exp(-p^2/\bar{p}^{2}_0 ) $$
 {\it vs.}
$$\lim_{y\to\infty}\eta_{J}(p) = \eta_{HOM}(p) =r^{3}_{0}\pi^{-3/2} \exp(-p^2/p^{2}_0 ).$$
Remind that here $\bar{r}_0 = \sqrt{3}/2 r_0$ and $\bar{p}_0 = \sqrt{3}/2 p_0 \ne \bar{r}_0^{-1}$ so we see one and the same renormalization of the parameters $r_0$ and $p_0$ in accordance with
the conclusion from \cite{ShePaMav06} that the so-called Tassie-Barker factors should be different for different distributions of particles in finite systems.

Now, one can ask to what extent the mean square radii of these DDs are modified due to the CMM corrections and the SRCs effects. The analytical expressions  of the FFs obtained here enable us to find an explicit
dependence of the corresponding radius on parameters $ r_{0}\, (p_{0}=r_{0}^{-1})$ and y.
In this connection, let us recall that it can be found  as coefficient of $-q^{2}/6$ in the conventional expression $F(q)=1-\frac{1}{6} q^2 r^{2}_{rms}+ \cdots $.
In particular, we get
$$
F_{J}(q)=1-\frac{1}{6} q^2 \langle r^{2}\rangle_{J}+ \cdots
$$
with
\begin{equation}
\langle r^{2}\rangle_{J} = -6 \frac{A^{\prime}_{J}(0)}{A_{J}(0)},
\label{equation124}
\end{equation}
where $A_{J}(q)$ is given by eq.(\ref{eq94}) (eq.(\ref{eq115})) in case of $^{4} He$ ($^{16} O$). Here $A^{\prime}_{J}(0)=\frac{d}{d q^{2}}A_{J}(q)\mid_{q=0}$.
Doing so, one can evaluate the difference $\Delta_{J}=\langle r^{2}\rangle_{J}-\langle r^{2}\rangle_{HOM}$, where $\langle r^{2}\rangle_{HOM} = \frac{3}{2} r_{0}^2\,\,
\left(\frac{9}{4} r_{0}^2\right)$ for $^{4}He (^{16}O)$. For example, $\Delta_{J}=0.282\, fm^2$ at $r_{0}=1.163\,fm $ and $y=3.120$ in case of $^{4}He$ and $\Delta_{J}=0.195 \,fm^2$ at $r_{0}=1.710\,fm$ and $y=20.192$ in case of $^{16}O$. It means that along with the aforementioned depression  the SRCs inclusion results in broadening the OBDD.

In its turn, the CMM correction contributes to
$$
F_{J,EST}(q)=1-\frac{1}{6}q^{2}\langle r^{2}\rangle_{J,EST}+ \cdots
$$
with
$$
\langle r^{2}\rangle_{J,EST}=c_{TB}+\langle r^{2}\rangle_{J},
$$
where $c_{TB}=-\langle r^2\rangle_{HOM}/A$. These quantities enter the expression
$$
\langle r^{2}\rangle_{CH}=c_{DW}+\langle r^{2}\rangle_{p}+\langle r^{2}\rangle_{J,EST}
$$
that determines the rms charge radii $\langle r^{2}\rangle_{CH} ^{1/2}$ to be extracted from
$$
F_{CH}(q)=1-\frac{1}{6}q^2 \langle r^{2}\rangle_{CH}+ \cdots.
$$
Remind their experimental values: $1.676 (2.730)fm $ for $^4 He\,(^{16}O)$, taken from \cite{deVries}.
One can verify that these values are reproduced by our calculations with
$\langle r^{2}\rangle_{J,EST}=2.000\,\, (6.644) fm^{2}$ for $^{4} He\,(^{16} O ))$ ($\langle r^{2}\rangle_{J}$ is equal to $2.31 fm^2$ and $6.77 fm^2$, respectively). Note also that accordingly the prescription \cite{ChaSau} $\langle r^{2}\rangle_{p}=0.775\,fm^2$.

\begin{figure}[h]
  \begin{center}
    \includegraphics[width=0.5\textwidth]{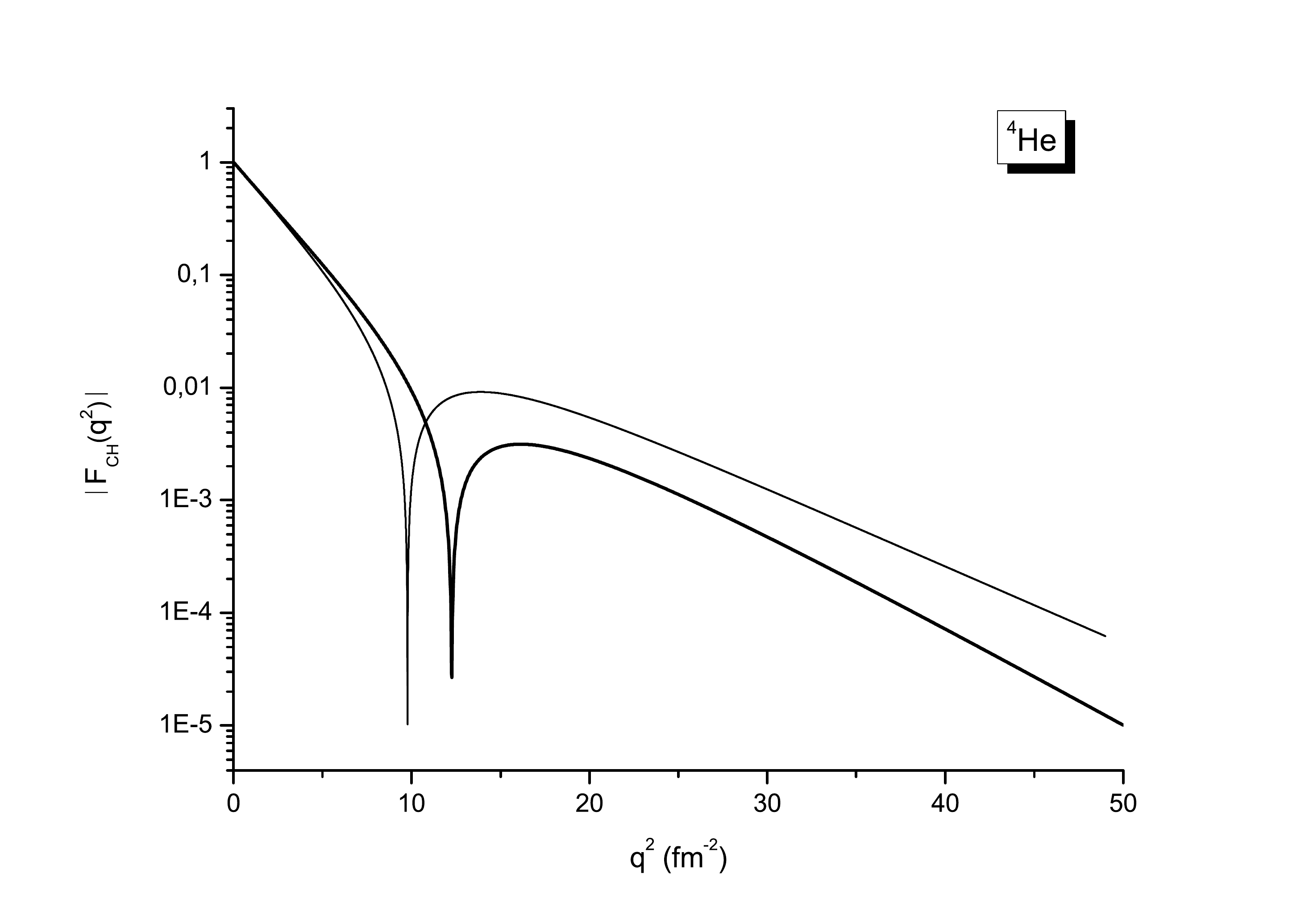}
  \end{center}
\caption{The exact point-like FF (thick solid line) for $^{4}He$ {\it vs} its approximation (solid line) by eq.(\ref{eq97})}
\end{figure}

Finally, one should note that we do not attach great importance to a fair agreement of our calculations with the data in figs. 1-2 and not too good one in fig. 5. In fact, as mentioned in sect.1, the IA, in which the charge FF is determined by formula  (\ref{FCH}), is insufficient (see, {\it e.g.}, \cite{KatAkaTan82}) to give an adequate treatment of the elastic electron scattering off nuclei with the q-increasing when  MEC effects become more and more important. In addition, one has to account for the higher-order contributions to the decompositions by eqs. (\ref{eq83})-(\ref{eq84}). Once more it illustrates fig.6, where we can see a considerable shift of the first diffraction minimum towards the larger $q$- values. Of course, the shift may be compensated by modifying  the values of the parameters involved.

Along with the pronounced flattening of the thick solid curve in fig. 5 in the vicinity of the $p=2fm^{-1}$ it means that every time higher-order correlations effects should be investigated separately (cf. similar results obtained in \cite{DalStringBoh82} for $^{4} He$ without any CM corrections). In the context, the large difference between the Argonne \cite{Argonne} and our calculations in fig. 5 at the $p$ - values $\gtrsim 2\,fm^{-1} $ can be explained to great extent by the inclusion of many-nucleon correlations in the former. Their role becomes stronger with increasing the mass number. In addition, being aware of the necessity \cite{Argonne} of introducing noncentral correlations (see also \cite{Alviol05}), we note that our method of restoring of the TI may be helpful for such complex numerical calculations as well.

\section{Summary}

We have shown how the approach developed in \cite{ShePaMav06} when studying the one-body and two-body density matrices of finite nuclei can be realized beyond the independent particle shell model. The appropriate treatment of the CMM is combined  with the inclusion of the SRCs in the nuclear WF, {\it e.g.}, regarding either the Jastrow ansatz or the UCOA. In our translationally invariant calculations the OBDD and OBMD are expectation values of the $A$-particle multiplicative operators which are dependent on the relative coordinates and momenta (the Jacobi variables) and sandwiched between intrinsic nuclear ground states.

An algebraic procedure proposed earlier helps us to avoid a cumbersome integration and see certain links between the distributions in question being expressed through one and the same generating function. In the course of the procedure the so-called Tassie-Barker factors stem directly from the intrinsic operators (not the WFs). One can stress that these factors being different, unlike other works (see, {\it e.g.}, \cite{Massen99} and \cite{Alviol05}), for the DD and MD occur by reflecting the translationally invariant structure of the corresponding intrinsic operators. Each of them is a Gaussian whose behavior in the space of variables is governed by the size parameter $r_{0}$(or its reciprocal $p_{0}$) and the particle number $A$ for a given finite system (nucleus), but it does not depend upon the choice of the g.s. WF. The latter can be a simple Slater determinant, include SRCs or not, be CMM-corrected or not, etc.

The use of the Cartesian or boson representation, in which the Jacobi variables are linear combinations of the creation {${\hat{\vec{a}}}^{\dag }$} and destruction {${\hat{\vec{a}}}$} operators for oscillator quanta, has allowed us to simplify the calculations for the closed shell nuclei $^{4} He$ and $^{16} O$. Certainly, the underlying idea based upon the normal ordering of the operators that meet the Bose commutation rules may be helpful in case of other closed and open shell nuclei. The analytic expressions for the intrinsic densities, form factors and momentum distributions derived in sect.2 with the Jastrow correlators are convenient in getting a deeper understanding of some nuclear properties. In particular, after restoring the TI on the SRCs background we have both in $\rho_{J}(r)$ and $\eta_{J}(p)$ their shrinking at enough large values of the ratio $y= \left(\frac{r_{0}}{r_{c}}\right)^{2}$.

Finally,regarding prospects of our approach in describing the interplay between the CMM and the SRC effects we mean, first of all, its application for calculations of the two-body momentum distributions in such reactions as $^4 He (e,e'NN)X$ and $^{16} O (e,e'NN)X$ (cf.the corresponding qualitative findings in \cite{ShePaMav06}). Our work in the subfield is in progress.

\appendix{}
\section{A key point of calculations beyond HOM}
\label{App:A.}
\setcounter{equation}{0}
\def\theequation{A.\arabic{equation}}
\hspace{0.5cm}

The algebraic technique, shown in sects. 2 and 3, can be also helpful in
calculating the expectations by eqs. (\ref{25}) and (\ref{28}) (or something like this) with WF  $\Phi$  that is either a linear
superposition of SDs or a SD which is composed of (HF) or other model orbitals
expanded in the HOM s.p. states. We find such expansions, {\it e.g.}, for HF
solutions \cite{GonInoKup77} and an effective inclusion \cite{RKB} of short-range
repulsion between nucleons (in both cases in spherical representation).

By definition, the normalized RKB-orbital (for a $1s^{4}$ configuration in $^{4}He$  nucleus) is
\begin{eqnarray}
 |\phi_{s}\rangle=\frac{1}{\sqrt{1+\beta^{2}}}\left(|\phi_{1s}\rangle+\beta|\phi_{2s}\rangle\right)
\label{A.1}
\end{eqnarray}
with an adjustable parameter $\beta$ .
In this connection, let us recall the well-known expressions for the HO orbitals $%
\mid nlm\rangle $ \ that are specified by the principal (spectroscopic),
orbital angular momentum and its projection quantum numbers $n$, $l$ and $m$. One has in coordinate space%
\begin{eqnarray}
 \varphi _{nlm}(\vec{r})=\left\langle \vec{r}\right\vert \left.
nlm\right\rangle =R_{nl}(r)Y_{lm}\left( \frac{\vec{r}}{r}\right)
\label{A.2}
\end{eqnarray}
\[
R_{nl}(r)=C_{nl}r_{0}^{-3/2}\left( \frac{r}{r_{0}}\right) ^{l}
\]
\[
\times\Phi \left(
1-n,l+\frac{3}{2};\frac{r^{2}}{r_{0}^{2}}\right) \exp \left( -\frac{1}{2}%
\frac{r^{2}}{r_{0}^{2}}\right) ,
\]%
\[
C_{nl}=\frac{\sqrt{2}}{\Gamma \left( l+\frac{3}{2}\right) }\left[ \frac{%
\Gamma \left( l+n+\frac{1}{2}\right) }{\Gamma (n)}\right] ^{1/2},
\]%
while in momentum space,%
\begin{eqnarray}
 \tilde{\varphi}_{nlm}(\vec{p})=\left\langle \vec{p}\right\vert \left.
nlm\right\rangle =\tilde{R}_{nl}(p)Y_{lm}\left( \frac{\vec{p}}{p}\right) ,
\label{A.3}
\end{eqnarray}
\[
\tilde{R}_{nl}(p)=(-1)^{n-1}(-i)^{l}C_{nl}p_{0}^{-3/2}\left( \frac{p}{p_{0}}%
\right) ^{l}\]
\[
\times\Phi \left( 1-n,l+\frac{3}{2};\frac{p^{2}}{p_{0}^{2}}\right)
\exp \left( -\frac{1}{2}\frac{p^{2}}{p_{0}^{2}}\right) ,
\]%
where following \cite{BaErd54}
$\Phi (a,c;x)$ is
the confluent function. By passing, remind also the link with the associated
Laguerre polynomials,%
\[
L_{n-1}^{l+1/2}(x)=\frac{\Gamma \left( l+n+\frac{1}{2}\right) }{\Gamma (l+%
\frac{3}{2})}\Phi \left( 1-n,l+\frac{3}{2};x\right) ,~n=1,2,....
\]

In turn, we find in the Cartesian representation
\begin{eqnarray}
|\varphi_{2s}\rangle=\sum_{n_{x}+n_{y}+n_{z}=2}|n_{x}n_{y}n_{z}\rangle\langle n_{x}n_{y}n_{z}|\varphi_{2s}\rangle
\label{A.4}
\end{eqnarray}
one can show (cf.,\cite{NS69})
\begin{eqnarray}
 |\varphi_{2s}\rangle=-\frac{1}{\sqrt{3}}\left( |200\rangle+|020\rangle+|002\rangle\right)
\label{A.5}
\end{eqnarray}
or taking in account eq.(\ref{e:bas}),
\begin{eqnarray}
|\varphi_{2s}\rangle=-\frac{1}{\sqrt{6}}{\hat{\vec{a}}}^{\dag }\cdot{\hat{\vec{a}}}^{\dag }|000\rangle,
\label{A.6}
\end{eqnarray}
i.e., for the RKB-orbital,
$$
|\phi_{s}\rangle=[1+{\beta}^2]^{-1/2}
[1-({\beta}/{\sqrt{6}})~\hat{\vec{a}}^{~\dag}\hat{\vec{a}}^{~\dag}~]\mid0\rangle.
\eqno{(A.7)}
$$
Substituting ({\rm A.7}) into ({\rm A.6}) (when calculating the
ratio $A^{IPM}(q)/A^{IPM}(0)$, the normalization factor $
[1+\beta^2]^{-1/2}$ can be omitted) we find
$$
\exp{(\vec{\chi}\cdot\vec{a})}\mid
\phi_{s} \rangle=[1-({\beta}/{\sqrt{6}})(\hat{\vec{a}}^{~\dag}+\vec{\chi})(\hat{\vec{a}}^{~\dag}+\vec{\chi})]\mid0\rangle
\eqno{(A.8)}
$$
for any complex vector $\vec{\chi}$.

Now, after modest effort we obtain
\[
\langle \phi_{s}\mid
\exp{(-\vec{\chi}~^*\cdot\hat{\vec{a}}^{~\dag})}\exp{(\vec{\chi}\cdot\hat{\vec{a}})}\mid
\phi_{s} \rangle=
\]
\[
=1+\beta^2-\frac{2}{3}~\beta^2\vec{\chi}~^*\vec{\chi}-
\]
$$
-\frac{\beta}{\sqrt{6}}~[\vec{\chi}~^*\vec{\chi}~^*+\vec{\chi}~\vec{\chi}~]+\frac{\beta^2}{6}~
(\vec{\chi}~^*\vec{\chi}~^*)(\vec{\chi}~\vec{\chi}) \eqno{(A.9)}
$$

\section{Relevant calculations}
\setcounter{equation}{0}
\def\theequation{B.\arabic{equation}}
\hspace{0.5cm}
The expectations of interest can be expressed in terms of these orbitals (in
general, the s.p. orbitals $\left\vert \lambda \right\rangle $ occupied in
the g.s.) in different ways. For example, using the formalism of secondary
quantization, one has%
\begin{eqnarray}
 A^{[2]}(q)&=\frac{1}{2A}Sp_{\sigma \tau }\sum_{\lambda _{1},\lambda _{2}\in
F}\langle \lambda _{1}\lambda _{2}\mid \hat{A}_{12}(\vec{q})\nonumber\\
&\times\mid \lambda
_{1}\lambda _{2}-\lambda _{2}\lambda _{1}\rangle ,
\label{B.1}
\end{eqnarray}
and %
\begin{eqnarray}
B^{[2]}(z)&=\frac{1}{2A}Sp_{\sigma \tau }\sum_{\lambda _{1},\lambda _{2}\in
F}\langle \lambda _{1}\lambda _{2}\mid \hat{B}_{12}(\vec{z})\nonumber\\
&\times\mid \lambda
_{1}\lambda _{2}-\lambda _{2}\lambda _{1}\rangle ,
\label{B.2}
\end{eqnarray}
where $F$ means the Fermi sea, so%
\begin{eqnarray}
 A^{[2]}(q)=A_{dir}^{[2]}(q)-A_{exc}^{[2]}(q),
\label{B.3}
\end{eqnarray}
\begin{eqnarray}
A_{dir}^{[2]}(q)=\frac{8}{A}\sum_{\lambda _{1},\lambda _{2}\in F}\langle
\varphi _{\lambda _{1}}\varphi _{\lambda _{2}}\mid \hat{A}_{12}(\vec{q})\mid
\varphi _{\lambda _{1}}\varphi _{\lambda _{2}}\rangle ,
\label{B.4}
\end{eqnarray}
\begin{eqnarray}
A_{exc}^{[2]}(q)=\frac{2}{A}\sum_{\lambda _{1},\lambda _{2}\in F}\langle
\varphi _{\lambda _{1}}\varphi _{\lambda _{2}}\mid \hat{A}_{12}(\vec{q})\mid
\varphi _{\lambda _{2}}\varphi _{\lambda _{1}}\rangle ,
\label{B.5}
\end{eqnarray}
and analogously%
\begin{eqnarray}
 B^{[2]}(z)=B_{dir}^{[2]}(z)-B_{exc}^{[2]}(z),
\label{B.6}
\end{eqnarray}
\begin{eqnarray}
 B_{dir}^{[2]}(z)=\frac{8}{A}\sum_{\lambda _{1},\lambda _{2}\in F}\langle
\varphi _{\lambda _{1}}\varphi _{\lambda _{2}}\mid \hat{B}_{12}(\vec{z})\mid
\varphi _{\lambda _{1}}\varphi _{\lambda _{2}}\rangle ,
\label{B.7}
\end{eqnarray}
\begin{eqnarray}
 B_{exc}^{[2]}(z)=\frac{2}{A}\sum_{\lambda _{1},\lambda _{2}\in F}\langle
\varphi _{\lambda _{1}}\varphi _{\lambda _{2}}\mid \hat{B}_{12}(\vec{z})\mid
\varphi _{\lambda _{2}}\varphi _{\lambda _{1}}\rangle .
\label{B.8}
\end{eqnarray}

We take the $ls$--coupling scheme with the orbitals
\begin{eqnarray}
\left\vert \lambda \right\rangle =\left\vert \varphi _{\lambda%
}\right\rangle \left\vert \chi _{\sigma \tau }\right\rangle .
 \label{B.11}
\end{eqnarray}
Accordingly eqs. (\ref{eq91}-\ref{eq92})%
\begin{eqnarray}
 &\hat{A}_{12}(\vec{q}) =\hat{h}^{\dagger }(1,2)\left[ e^{i\vec{q}\widehat{%
\vec{r}}_{1}}+e^{i\vec{q}\widehat{\vec{r}}_{2}}\right] \hat{h}(1,2)\nonumber\\
&+\hat{h}%
^{\dagger }(1,2)\left[ e^{i\vec{q}\widehat{\vec{r}}_{1}}+e^{i\vec{q}\widehat{%
\vec{r}}_{2}}\right]\nonumber\\
&+\left[ e^{i\vec{q}\widehat{\vec{r}}_{1}}+e^{i\vec{q}\widehat{\vec{r}}_{2}}%
\right] \hat{h}(1,2),
\label{B.9}
\end{eqnarray}
\begin{eqnarray}
 &\hat{B}_{12}(\vec{z}) =\hat{h}^{\dagger }(1,2)\left[ e^{i\vec{z}\widehat{%
\vec{p}}_{1}}\nonumber+e^{i\vec{z}\widehat{\vec{p}}_{2}}\right] \hat{h}(1,2)
\nonumber\\
&+\hat{h}%
^{\dagger }(1,2)\left[ e^{i\vec{z}\widehat{\vec{p}}_{1}}+e^{i\vec{z}\widehat{%
\vec{p}}_{2}}\right]
\nonumber\\
&+\left[ e^{i\vec{z}\widehat{\vec{p}}_{1}}+e^{i\vec{z}\widehat{\vec{p}}_{2}}%
\right] \hat{h}(1,2),
\label{B.10}
\end{eqnarray}
once $\hat{f}(\alpha ,\beta )=1+\hat{h}(\alpha ,\beta )$~($\alpha ,\beta
=1,...,A$). In this work calculations have been carried out with the
state-independent correlator%
\begin{eqnarray}
 \hat{h}(\alpha ,\beta )=h\left( \left\vert \widehat{\vec{r}}_{\alpha }-%
\widehat{\vec{r}}_{\beta }\right\vert \right) =-\exp \left[ -\frac{\left(
\widehat{\vec{r}}_{\alpha }-\widehat{\vec{r}}_{\beta }\right) ^{2}}{r_{c}^{2}%
}\right],\,\nonumber\\
\label{B.11}
\end{eqnarray}
where $r_{c}$ is a correlation radius.

Further, putting in the relation (cf. eq. \ref{eq63}),%
\[
\exp \left[ -\vec{u}^{\ast }\widehat{\vec{a}}^{\dagger }+\vec{u}\widehat{%
\vec{a}}\right] =e^{-\frac{1}{2}\vec{u}^{\ast }\vec{u}}\exp \left[ -\vec{u}%
^{\ast }\widehat{\vec{a}}^{\dagger }\right] \exp \left( \vec{u}\widehat{\vec{%
a}}\right)
\]%
the vector $\vec{u}$ equal first to%
\begin{eqnarray}
 \vec{u}=i\frac{r_{0}}{\sqrt{2}}\vec{q}
\label{B.12}
\end{eqnarray}
and second to%
\begin{eqnarray}
 \vec{u}=\frac{p_{0}}{\sqrt{2}}\vec{z}
\label{B.13}
\end{eqnarray}
we split exponents $\exp \left( i\vec{q}\widehat{\vec{r}}\right) $ and
$\exp\left( i\vec{z}\widehat{\vec{p}}\right) $, respectively, in eqs. (\ref{B.9}) and (\ref{B.10}) into
such a normally ordered form. Then, when evaluating the sums in eqs.
(\ref{B.4})--(\ref{B.5}) ((\ref{B.7})--(\ref{B.8})), it suffices to consider the matrix elements:%
\begin{eqnarray}
 &M_{\lambda _{1}\lambda _{2}}^{(k)}(\vec{u})=\langle \varphi _{\lambda
_{1}}\varphi _{\lambda _{2}}\mid e^{-\vec{u}^{\ast }\widehat{\vec{a}}%
_{1}^{\dagger }}H^{(k)}(\widehat{\vec{r}};\vec{u})\nonumber\\
&\times e^{\vec{u}\widehat{\vec{a}}%
_{1}}\mid \varphi _{\lambda _{1}}\varphi _{\lambda _{2}}\rangle ,
\label{B.14}
\end{eqnarray}
\begin{eqnarray}
 &\bar{M}_{\lambda _{1}\lambda _{2}}^{(k)}(\vec{u})=\langle \varphi _{\lambda
_{1}}\varphi _{\lambda _{2}}\mid e^{-\vec{u}^{\ast }\widehat{\vec{a}}%
_{1}^{\dagger }}H^{(k)}(\widehat{\vec{r}};\vec{u})\nonumber\\
&\times e^{\vec{u}\widehat{\vec{a}}%
_{1}}\mid \varphi _{\lambda _{2}}\varphi _{\lambda _{1}}\rangle
(k=1,2),
\label{B.15}
\end{eqnarray}
we have employed the property (\ref{44}) and introduced the operators%
\begin{eqnarray}
 H^{(1)}(\widehat{\vec{r}};\vec{u})=h\left( \left\vert \widehat{\vec{r}}+%
\frac{r_{0}}{\sqrt{2}}\vec{u}\right\vert \right)
\label{B.16}
\end{eqnarray}
and%
\begin{eqnarray}
 H^{(2)}(\widehat{\vec{r}};\vec{u})=H^{(1)}(\widehat{\vec{r}};-\vec{u}^{\ast
})H^{(1)}(\widehat{\vec{r}};\vec{u})
\label{B.17}
\end{eqnarray}
dependent on the distance $\vec{r}=\vec{r}_{1}-\vec{r}_{2}$ between the
nucleons. Obviously, the superscript $k$ in $H^{(k)}(\widehat{\vec{r}};\vec{u%
})$ labels the order in the correlations involved.

Using the definition (\ref{A.2}) and the transformation (\ref{eq113}) the contributions of
interest can be represented as%
\begin{eqnarray}
 \pi ^{3}\sum_{m}M_{1s;1pm}^{(k)}(\vec{u})=r_{0}^{-6}\int d\vec{r}_{1}\int d%
\vec{r}_{2}\frac{r_{1}^{2}+r_{2}^{2}}{r_{0}^{2}}\nonumber\\
\times e^{-\frac{r_{1}^{2}+r_{2}^{2}%
}{r_{0}^{2}}}H^{(k)}(\vec{r};\vec{u})\equiv I^{(k)}(u),
\label{B.18}
\end{eqnarray}
\begin{eqnarray}
&\pi ^{3}\sum_{m}M_{1pm;1s}^{(k)}(\vec{u})=r_{0}^{-6}\int d\vec{r}_{1}\int d%
\vec{r}_{2}e^{-\frac{r_{1}^{2}+r_{2}^{2}}{r_{0}^{2}}}H^{(k)}(\vec{r}%
;\vec{u})\nonumber\\
&\times \left( \frac{r_{1}^{2}+r_{2}^{2}}{r_{0}^{2}}-\vec{u}^{\ast }\vec{u}+%
\frac{\vec{u}-\vec{u}^{\ast }}{\sqrt{2}}\frac{\vec{r}_{1}-\vec{r}_{2}}{r_{0}}%
\right) \equiv J^{(k)}(u)\nonumber\\
\label{B.19}
\end{eqnarray}
\begin{equation}
\begin{split}
\pi ^{3}&M_{pp}^{(k)}(\vec{u})=r_{0}^{-6}\int d\vec{r}_{1}\int d\vec{r}%
_{2}e^{-\frac{r_{1}^{2}+r_{2}^{2}}{r_{0}^{2}}}\\
&H^{(k)}(\vec{r};\vec{%
u})\times 4\{ \frac{r_{1}^{2}r_{2}^{2}}{r_{0}^{4}}-\frac{%
r_{1}^{2}+r_{2}^{2}}{4r_{0}^{2}}\vec{u}^{\ast }\vec{u}\\
+&\frac{\vec{u}-\vec{u}%
^{\ast }}{2\sqrt{2}r_{0}^{3}}\left[ \left( r_{1}^{2}+r_{2}^{2}\right) \left(
\vec{r}_{1}-\vec{r}_{2}\right)  -r_{1}^{2}\vec{r}_{1}+r_{2}^{2}\vec{r}_{2}%
\right] \} \equiv P^{(k)}(u),\\
&\,\,\,\,\,\,\,(k=1,2),
\label{B.20}
\end{split}
\end{equation}
while
\begin{equation}
\begin{split}
&\pi ^{3}\sum_{m}\bar{M}_{1s;1pm}^{(k)}(\vec{u})=2r_{0}^{-6}\int d\vec{r}%
_{1}\int d\vec{r}_{2}\\
&\times\left[ \frac{\vec{r}_{1}}{r_{0}}+\frac{\vec{u}}{\sqrt{2}%
}\right]\frac{\vec{r}_{2}}{r_{0}}e^{-\frac{r_{1}^{2}+r_{2}^{2}}{r_{0}^{2}}%
}H^{(k)}(\vec{r};\vec{u})\equiv \bar{I}^{(k)}(u),
\label{B.21}
\end{split}
\end{equation}
\begin{equation}
\begin{split}
&\pi ^{3}\sum_{m}\bar{M}_{1pm;1s}^{(k)}(\vec{u})=2r_{0}^{-6}\int d\vec{r}%
_{1}\int d\vec{r}_{2}\\
&\times\left[ \frac{\vec{r}_{1}}{r_{0}}-\frac{\vec{u}^{\ast }}{%
\sqrt{2}}\right]  \times\frac{\vec{r}_{2}}{r_{0}}e^{-\frac{r_{1}^{2}+r_{2}^{2}}{%
r_{0}^{2}}}H^{(k)}(\vec{r};\vec{u})\equiv \bar{J}^{(k)}(u),
\label{B.22}
\end{split}
\end{equation}
\begin{equation}
\begin{split}
\pi ^{3}&\sum_{m_{1}m_{2}}\bar{M}_{1pm_{1};1pm_{2}}^{(k)}(\vec{u}%
)=4r_{0}^{-6}\int d\vec{r}_{1}\int d\vec{r}_{2}e^{-\frac{r_{1}^{2}+r_{2}^{2}}{r_{0}^{2}}}\\
&\times H^{(k)}(\vec{r};\vec{u})\{ \frac{(\vec{r_1}\vec{r_2})^{2}}{r_0^{4}}- \frac{(\vec{r_1}\vec{r_2})%
(\vec{u}^{*}\vec{r_2})}{\sqrt{2}r_0^{3}}\\
&+\frac{(\vec{r_1}\vec{r_2})(\vec{u}\vec{r_2})}{\sqrt{2}r_0^{3}}- \frac{(\vec{u}\vec{r_2})(\vec{u^{*}}\vec{r_2})}{2r_0^{2}}\}%
\equiv \bar{P}^{(k)}(u),\\
&k=1,2.
\label{B.23}
\end{split}
\end{equation}
Substituting expressions (\ref{B.16})--(\ref{B.17}) into these equations, we find with
the correlator (\ref{B.11}),%
\begin{eqnarray}
 &e^{\frac{1}{2}yu^{2}}I^{(1)}(u)=-\int d\vec{r}_{1}\int d\vec{r}%
_{2}e^{-r_{1}^{2}-r_{2}^{2}}e^{-yr^{2}}\nonumber\\
&e^{-\sqrt{2}y\vec{u}\vec{r}}\left( r_{1}^{2}+r_{2}^{2}\right),
\label{B.24}
\end{eqnarray}
\begin{equation}
\begin{split}
&e^{\frac{1}{2}yu^{2}}J^{(1)}(u)=-\int d\vec{r}_{1}\int d\vec{r}%
_{2}e^{-r_{1}^{2}-r_{2}^{2}}e^{-yr^{2}}e^{-\sqrt{2}y\vec{u}\vec{r}}\\
&\times ( r_{1}^{2}+r_{2}^{2}-\vec{u}^{\ast }\vec{u}+\frac{\vec{u}-\vec{u%
}^{\ast }}{\sqrt{2}}\vec{r}),
\label{B.25}
\end{split}
\end{equation}
\begin{equation}
\begin{split}
&e^{\frac{1}{2}yu^{2}}P^{(1)}(u)=-\int d\vec{r}_{1}\int d\vec{r}%
_{2}e^{-r_{1}^{2}-r_{2}^{2}}e^{-yr^{2}}e^{-\sqrt{2}y\vec{u}\vec{r}}\\
& \times 4\{ r_{1}^{2}r_{2}^{2}-\frac{r_{1}^{2}+r_{2}^{2}}{4}\vec{u}^{\ast
}\vec{u}\\
&+\frac{\vec{u}-\vec{u}^{\ast }}{2\sqrt{2}}\left[ \left(
r_{1}^{2}+r_{2}^{2}\right) \vec{r}-r_{1}^{2}\vec{r}_{1}+r_{2}^{2}\vec{r}_{2}%
\right] \} ,
\label{B.26}
\end{split}
\end{equation}
and%
\begin{eqnarray}
& e^{\frac{1}{2}yu^{2}}\bar{I}^{(1)}(u)=-2\int d\vec{r}_{1}\int d\vec{r}%
_{2}e^{-r_{1}^{2}-r_{2}^{2}}e^{-yr^{2}}e^{-\sqrt{2}y\vec{u}\vec{r}}\nonumber\\
&\times\left[
\vec{r}_{1}+\frac{\vec{u}}{\sqrt{2}}\right] \vec{r}_{2},
\label{B.27}
\end{eqnarray}
\begin{eqnarray}
 &e^{\frac{1}{2}yu^{2}}\bar{J}^{(1)}(u)=-2\int d\vec{r}_{1}\int d\vec{r}%
_{2}e^{-r_{1}^{2}-r_{2}^{2}}e^{-yr^{2}}e^{-\sqrt{2}y\vec{u}\vec{r}}\nonumber\\
&\times\left[
\vec{r}_{1}-\frac{\vec{u}^{\ast }}{\sqrt{2}}\right] \vec{r}_{2},
\label{B.28}
\end{eqnarray}
\begin{eqnarray}
&e^{\frac{1}{2}yu^{2}}\bar{P}^{(1)}(u)=-4\int d\vec{r}_{1}\int d\vec{r}%
_{2}e^{-r_{1}^{2}-r_{2}^{2}}e^{-yr^{2}}e^{-\sqrt{2}y\vec{u}\vec{r}}\nonumber\\
&\times[ \frac{(\vec{r_1}\vec{r_2})^{2}}{r_0^{4}}- \frac{(\vec{r_1}\vec{r_2})%
(\vec{u}^{*}\vec{r_2})}{\sqrt{2}r_0^{3}}\nonumber\\
&+\frac{(\vec{r_1}\vec{r_2})(\vec{u}\vec{r_2})}{\sqrt{2}r_0^{3}}- \frac{(\vec{u}\vec{r_2})(\vec{u^{*}}\vec{r_2})}{2r_0^{2}}
].
\label{B.29}
\end{eqnarray}

It is readily seen that the corresponding counterparts of the second order,
multiplied by the same factor $-\exp \left[ -\frac{1}{2}y\left( \vec{u}%
^{\ast 2}+\vec{u}^{2}\right) \right] $, can be obtained from the integrals
(\ref{B.24})--(\ref{B.26}) and (\ref{B.27})--(\ref{B.29}) by doing in their integrands the two
independent changes: $y\rightarrow 2y$ and $y\vec{u}\rightarrow y\left( \vec{%
u}-\vec{u}^{\ast }\right) $. In turn, these integrals may be calculated by
addressing an auxiliary integral%
\begin{eqnarray}
&I\left( \vec{u};a,\vec{b}_{1},\vec{b}_{2}\right) =\int d\vec{r}_{1}\int d%
\vec{r}_{2}e^{-a\left( r_{1}^{2}+r_{2}^{2}\right) }\nonumber\\
& e^{-yr^{2}}e^{-\sqrt{2}y%
\vec{u}\vec{r}}\exp \left[ \vec{b}_{1}\vec{r}_{1}+\vec{b}_{2}\vec{r}_{2}%
\right]\nonumber
\end{eqnarray}
in the vicinity of the parameter values: $a=1$, $\vec{b}_{1}=\vec{b}_{2}=0$.
Indeed, we have
\begin{eqnarray}
 I\left( \vec{u};a,\vec{b}_{1},\vec{b}_{2}\right) =\frac{\pi ^{3}}{\left[
a\left( a+2y\right) \right] ^{3/2}}\nonumber\\
\times\exp \left[ \frac{\vec{B}^{2}}{8a}+\frac{%
\left( \vec{b}-\sqrt{2}y\vec{u}\right) ^{2}}{2a+4y}\right],\label{B.30}
\end{eqnarray}
where $\vec{B}=\vec{b}_{1}+\vec{b}_{2}$ and $\vec{b}=\frac{1}{2}\left( \vec{b%
}_{1}-\vec{b}_{2}\right), $ and after evident differentiating (for instance, using analytic means of \emph{Mathematica})
we get formulae (eq.\ref{eq114})-(eq.\ref{eq119}).

It yields
\begin{equation}
\left\{
\begin{array}{c}
A_{dir}^{[2]}(q) \\
B_{dir}^{[2]}(z)%
\end{array}%
\right\} =e^{-\frac{1}{2}\vec{u}^{\ast }\vec{u}}\left[ M_{dir}^{(2)}(\vec{u}%
)+2
Re\,M_{dir}^{(1)}(\vec{u})\right]
\label{B31}
\end{equation}%
and%
\begin{equation}
\left\{
\begin{array}{c}
A_{exc}^{[2]}(q) \\
B_{exc}^{[2]}(z)%
\end{array}%
\right\} =\frac{1}{4}e^{-\frac{1}{2}\vec{u}^{\ast }\vec{u}}\left[
M_{exc}^{(2)}(\vec{u})+2
Re\, M_{exc}^{(1)}(\vec{u})\right]
\label{eq121}
\end{equation}%
where the argument $\vec{u}=i\frac{r_{0}}{\sqrt{2}}\vec{q}$ ( $\vec{u}=\frac{p_{0}}{\sqrt{2}}\vec{z}$) for {A}({B}),
with
\begin{equation}
M_{dir}^{(k)}(\vec{u})=\sum_{\chi _{1},\lambda _{2}\in F}M_{\lambda
_{1}\lambda _{2}}^{(k)}(\vec{u})  \label{eq122}
\end{equation}%
and%
\begin{equation}
M_{exc}^{(k)}(\vec{u})=\sum_{\chi _{1},\lambda _{2}\in F}\bar{M}_{\lambda
_{1}\lambda _{2}}^{(k)}(\vec{u}),~(k=1,2).  \label{eq123}
\end{equation}%
Now we will separate out the purely $1s$ subshell, mixed $1s-1p$ and purely $%
1p$ subshell contributions assuming%
\begin{equation}
M_{dir}^{(k)}(\vec{u})=M_{ss}^{(k)}(\vec{u})+M_{mix}^{(k)}(\vec{u}%
)+M_{pp}^{(k)}(\vec{u})  \label{eq124}
\end{equation}%
and%
\begin{equation}
M_{exc}^{(k)}(\vec{u})=\bar{M}_{ss}^{(k)}(\vec{u})+\bar{M}_{mix}^{(k)}(\vec{u%
})+\bar{M}_{pp}^{(k)}(\vec{u}),~(k=1,2)  \label{eq125}
\end{equation}%
with%
\begin{equation}
M_{ss}^{(k)}(\vec{u})=\bar{M}_{ss}^{(k)}(\vec{u})=\langle 1s1s\mid H^{(k)}(%
\widehat{\vec{r}};\vec{u})\mid 1s1s\rangle ,  \label{eq126}
\end{equation}%
\begin{equation}
M_{mix}^{(k)}(\vec{u})=\sum_{m}\left[ M_{1s;1pm}^{(k)}(\vec{u}%
)+M_{1pm;1s}^{(k)}(\vec{u})\right] ,  \label{eq127}
\end{equation}%
\begin{equation}
M_{pp}^{(k)}(\vec{u})=\sum_{m_{1}m_{2}}M_{1pm_{1};1pm_{2}}^{(k)}(\vec{u}),
\label{eq128}
\end{equation}%
and analogously for the bar quantities.

%
%

\end{document}